\documentclass[journal,twoside,web]{ieeecolor}
\usepackage{generic,cite}
\usepackage{amsmath,amssymb,amsfonts, mathtools}
\usepackage{algorithm, algpseudocode}
\usepackage[pdftex]{graphicx}
\usepackage{textcomp, subcaption}

\usepackage{soul}

\newtheorem{theorem}{Theorem}
\newtheorem{lemma}{Lemma}
\newtheorem{proposition}{Proposition}

\newtheorem{assumption}{Assumption}

\newtheorem{remark}{Remark}

\usepackage[table]{xcolor}
\newcommand{\hsp}{\hspace{1pt}}

\usepackage{bigstrut, xfrac, subfiles}
\def\BibTeX{{\rm B\kern-.05em{\sc i\kern-.025em b}\kern-.08em
T\kern-.1667em\lower.7ex\hbox{E}\kern-.125emX}}
\begin{document}
\title{
Distributed Error-Identification and Correction using Block-Sparse Optimization}
\author{
Shiraz Khan and Inseok Hwang, \IEEEmembership{Member, IEEE}
\thanks{\textcolor{blue}{© 2023 IEEE. Personal use of this material is permitted. Permission from IEEE must be obtained for all other uses, in any current or future media, including reprinting/republishing this material for advertising or promotional purposes, creating new collective works, for resale or redistribution to servers or lists, or reuse of any copyrighted component of this work in other works.}}
\thanks{The authors are with the School of Aeronautics and Astronautics, Purdue University,
West Lafayette, IN 47906. Email: {\tt\small shiraz, ihwang@purdue.edu}}
\thanks{This research is funded by the Secure Systems Research Center (SSRC) at the Technology Innovation Institute (TII), UAE. The authors are grateful to Dr. Shreekant (Ticky) Thakkar and his team members at the SSRC for their valuable comments and support.}
}
\maketitle
\begin{abstract}
The conventional solutions for fault-detection, identification, and reconstruction (FDIR) require centralized decision-making mechanisms which
are typically combinatorial in their nature, necessitating the design of an efficient distributed FDIR mechanism that is suitable for multi-agent applications.
To this end, we develop a general framework for efficiently reconstructing a sparse vector being observed over a sensor network via nonlinear measurements. The proposed framework is used to design a distributed multi-agent FDIR algorithm based on a combination of the sequential convex programming (SCP) and the alternating direction method of multipliers (ADMM) optimization approaches. The proposed distributed FDIR algorithm can process a variety of inter-agent measurements (including distances, bearings, relative velocities, and subtended angles between agents) to identify the faulty agents and recover their true states.
The effectiveness of the proposed distributed multi-agent FDIR approach is demonstrated by considering a numerical example in which the inter-agent distances are used to identify the faulty agents in a multi-agent configuration, as well as reconstruct their error vectors.
\end{abstract}
%
%
\section{Introduction}
\label{sec:introduction}
A Cyber-Physical System (CPS) is an autonomous agent that has a physical state, as well as onboard computers and sensors that enable it to observe and navigate its environment safely.
Two or more CPS can be collectively deployed as a multi-agent system to achieve a series of objectives, such that the capabilities of the multi-agent system far exceed those of the individual CPS. 
However, the increased capabilities of multi-agent systems go hand in hand with their increased complexity, making the design of efficient estimation and decision-making algorithms for multi-agent systems a challenging task. This is especially true in the case of algorithms that have a combinatorial aspect to them, such as the conventional algorithms for fault-detection, identification, and reconstruction (FDIR)\footnote{In the literature, the acronym FDIR is sometimes expanded as fault diagnosis, isolation, and recovery \cite{naderi2017datadriven_FDIR, guo2012distributed}.}. Fault detection refers to the mechanism by which an agent's sensor data is monitored for the presence of systematic anomalies
that manifest as a discrepancy between the estimated state and the true state of the agent. We refer to the presence of such a discrepancy at a given agent as an \textit{error}. In practice, errors can arise due to various types of sensor or actuator faults, including sensor bias, miscalibration, cyberattacks, modeling errors, and programming bugs \cite{gao2022survey}. 
The objective of multi-agent FDIR is then to identify the agents which have faults and correct their error vectors, respectively. In particular, the multi-agent FDIR problem does not discriminate between the types of faults at the agent level, but rather focuses on fault-identification at the network level, which involves identifying all the agents in the network that have faults (see, for example, \cite{kazumune2020distributed}).

Historically, FDIR algorithms were first developed for the single-agent, multi-sensor scenario. As single-agent systems are lower in dimension and have fewer bottlenecks than multi-agent systems, the FDIR problem can be solved using data-driven methods \cite{naderi2017datadriven_FDIR}, banks of observers \cite{hwang2009survey},  or
combinatorial algorithms like sensor subset search \cite{mishra2016secure}. In general, the FDIR mechanisms designed for single-agent systems are not feasible for use in multi-agent systems due to the collective state vector of a multi-agent system being relatively high-dimensional.
Moreover, the agents are typically limited in their communication and computational capabilities, which rules out the possibility of processing the measurement information in a centralized manner. 
Therefore, several authors have proposed alternative FDIR strategies that are better suited for multi-agent applications, including those based on $H_\infty$ performance indices \cite{chadli2017distributed_hinf_iet, gallehdari2017h}, residual testing \cite{guo2012distributed}, $l_1$ norm minimization \cite{kazumune2020distributed}, and interval observers \cite{zhang2018distributed} among others \cite{zhang2021physical_survey}. 
A limitation of the preceding works is that they do not accommodate the nonlinear measurement models that typically arise in real-world CPS applications, such as inter-agent distance and bearing measurements,
which present unique challenges to the multi-agent FDIR problem. 
For instance, it is known that the inter-agent distances are not sufficient to uniquely determine the identities and states of the faulty agents, which means that additional information channels, assumptions, or regularization techniques are needed in order to solve the FDIR problem \cite{khan2023recovery}. 

The study of the observability and controllability of multi-agent systems which are able to obtain nonlinear inter-agent measurements, including distance \cite{oh2015survey,topology_const2015observability}, bearing \cite{zhao2019bearing}, and subtended angle \cite{weak_rigidity} measurements, is collectively referred to as \textit{rigidity theory}. In the literature on rigidity theory, it is assumed that some of the agents, called the anchors, are able to observe their true states, and that the identities of the anchors are known to all the agents \textit{a priori} \cite{zhao2019bearing}. Under this assumption, the agents that are classified as the anchors are exempted from having faults, which ensures that the faults in the remaining agents can be corrected by processing the inter-agent measurements.
However, under the anchor-based approach to multi-agent FDIR, the presence of undetected faults at the anchors can cause the FDIR algorithm to misidentify the faulty agents, since the anchors were assumed to be fault-free. This fact is especially problematic when the faults are introduced in an adversarial manner, e.g., by a cyberattacker who perturbs some of the anchors' state estimates using sensor spoofing attacks, thereby compromising the FDIR capabilities of the entire network.

To address these limitations of existing multi-agent FDIR mechanisms, we propose a novel framework for multi-agent FDIR that does not require any of the agents to be classified as the anchors, and can therefore diagnose the entire multi-agent network for faults. In the proposed framework, the multi-agent FDIR problem is reformulated as one of reconstructing a block-sparse error vector. Thereafter, the nonlinearity of the inter-agent measurements is accommodated using sequential convex programming (SCP) \cite{scp_zillober2004}, and a distributed implementation of the error-reconstruction algorithm is developed using the alternating direction method of multipliers (ADMM) optimization procedure \cite{boyd2011distributed}.
Once the block-sparse multi-agent error vector is reconstructed,
the presence of non-zero blocks in the reconstructed error vector indicates the presence of faults, the positions of the non-zero blocks identify the faulty agents, and the non-zero blocks are precisely the error vectors of the faulty agents; thus, the objectives of multi-agent FDIR are accomplished in an efficient, distributed manner. A secondary goal of this paper is to explain why the proposed approach works, and how it relates to the existing literature on rigidity theory. 

In summary, the contributions of this work are as follows:
\begin{enumerate}
    \item The multi-agent FDIR problem is reformulated as an optimization problem that searches for a block-sparse vector subject to nonlinear equality constraints.
    \item The search-space (i.e., the feasible set) of the preceding problem is characterized using tools from differential geometry. It is shown that our characterization generalizes (and moreover, reinterprets) some of the existing results in rigidity theory.
    \item A combination of the SCP and ADMM optimization techniques is used to develop a distributed algorithm for reconstructing the error vector, which exploits the unique structure of the multi-agent FDIR problem.
    \item The theory of subgradient methods is used to show that the proposed algorithm exhibits a \textit{thresholding property}, by which each agent compares the norm of a residual vector against a pre-specified threshold in order to detect and identify the faults.
    \item Finally, a numerical example of multi-agent FDIR using inter-agent distance measurements is used to demonstrate the effectiveness of the proposed approach.
\end{enumerate}

The remainder of this paper is organized as follows. Section \ref{sec:prob_form} presents some preliminary definitions and assumptions, and formulates the multi-agent FDIR problem. In Section \ref{sec:reconstruction}, we characterize the search-space of the multi-agent FDIR problem, explain its connections to rigidity theory, and motivate the assumption of block-sparsity of the error vector. In Section \ref{sec:distributed}, the distributed multi-agent FDIR algorithm is developed and analyzed. Section \ref{sec:numerical} presents numerical simulations that validate our theoretical results and demonstrate the effectiveness of the proposed algorithm. Finally, the conclusions and future research directions are discussed in Section \ref{sec:conclusions}.
\section{Problem Formulation}
\label{sec:prob_form}
\subsection{Notational Preliminaries}
\label{subsec:notation}

Let $n_1, n_2, \dots, n_k$ be a sequence of $k$ positive integers whose sum is denoted by $n \coloneqq \sum_{i=1}^k n_i$.
Given a block vector $\mathbf v\in\mathbb R^{n}$ that is partitioned into $k$ blocks of lengths $n_1, n_2, \dots, n_k$, $\mathbf v[i]\in \mathbb R^{n_i}$ refers to the $i^{th}$ block of $\mathbf v$. Similarly, given a block matrix $\mathbf A$, its $(i,j)^{th}$ block (whose dimensions are understood from context) is denoted as $\mathbf A[i,j]$. The $l_2$ norm (or the Euclidean norm) of $\mathbf v$ is denoted by $\|\mathbf v\|$.
Similar to \cite{efficient_block_sparse_2010, robust_NSP_2017}, we define the following notation:
\[
\|\mathbf v\|_{2,q}=
\begin{array}{lc}
     \Big( \sum_{i=1}^{k} \|\mathbf v[\small i]\|^q\Big)^{1/q} 
\end{array}
\]
where $0<q<\infty$. It follows from the definition that $\|\mathbf v\|_{2,2} = \|\mathbf v\|$.
The case of $q=0$ can be evaluated using limits, giving us the definition,
\[
\|\mathbf v\|_{2,0} = \sum_{i=1}^{k} \mathbb I\Big(\|\mathbf v[\small i]\|>0\Big)
\]
where $\mathbb I(\hspace{1pt}\cdot\hspace{1pt})$ is the indicator function which is equal to $1$ when the inequality holds,
and $0$ otherwise. 
Given a set $\mathcal S$, $|\mathcal S|$ denotes its cardinality.
Given an index set $\mathcal S\subset \{1, 2, \dots, k\}$, $\mathcal S^\complement$ denotes its complement, given by $\{1, 2, \dots, k\}\backslash \mathcal S$. 
\subsection{Sensing and Communication Topology}
\label{subsec:graphs}
An (undirected) \textit{hypergraph} refers to a pair $(\mathcal V, \mathcal E)$ constituting a set of vertices $\mathcal V$ and a set of hyperedges $\mathcal E$, where each element of $\mathcal E$ (called a hyperedge) is a subset of $\mathcal V$ \cite[Sec. 1.10]{diestel2017}. Consider a multi-agent system represented as a hypergraph, $\mathcal G=(\mathcal V, \mathcal E)$, such that the vertices correspond to the agents (of which there are $|\mathcal V|$ in total), and each hyperedge represents the availability of a measurement that depends on the states of the corresponding agents. As hypergraphs generalize graphs by allowing more than two (or even zero or one) vertices to be connected, they enable us to represent measurements that are functions of three or more agents' states, such as subtended angle and time-difference-of-arrival (TDoA) measurements \cite{weak_rigidity,tdoa_2017}. Let $\mathcal V$ and $\mathcal E$ be endowed with arbitrary orderings, so that we may write $\mathcal V = \lbrace 1, 2, \dots, |\mathcal V|\rbrace$ and $\mathcal E= \lbrace \mathcal E^{(1)}, \mathcal E^{(2)}, \dots, \mathcal E^{(|\mathcal E|} \rbrace$.
Agents $i$ and $j$ are said to be \textit{neighbors} of each other if, for some $l\in\lbrace 1, \dots, |\mathcal E|\rbrace$, both $i$ and $j$ are in $\mathcal E^{(l)}$. The set of neighbors of agent $i$ is denoted by $\mathcal N_i \subseteq \mathcal V$.
Each agent in $\mathcal G$ is assumed to be a Cyber-Physical System (CPS) that has a physical state, an embedded computer, and the capabilities to communicate and (potentially) obtain measurements through a variety of sensors. We make the following assumption about the communication capabilities of the agents.

\vspace{2pt}
\begin{assumption}[Communication Topology]
Each agent is able to communicate with its neighbors, and the agents are able to synchronize\footnote{The meaning of `synchronize' as it is used in Assumption \ref{ass:1} can be understood unambiguously by studying the algorithm which we develop in Section \ref{sec:distributed}.} their communications.
\label{ass:1}
\end{assumption}
\vspace{2pt}


Thus, we have implicitly made the assumption that if agent $i$ is able to sense agent $j$, then agents $i$ and $j$ are able to establish a bidirectional communication channel between them as well. 

\subsection{Agent States and Estimates}
The collective state of the multi-agent system is represented by a block vector $\mathbf p$, which has the form 
\begin{equation}
\mathbf p = \begin{bmatrix} \mathbf p[1]^\top & \mathbf p [2]^\top & \dots & \mathbf p [|\mathcal V|]^\top \end{bmatrix}^\top \in \mathbb R^n
\end{equation}
where $\mathbf p[i] \in \mathbb R^{n_i}$ is the $i^{th}$ agent's state, $n_1, n_2, \dots, n_{|\mathcal V|}$ are positive integers representing the dimensions of each of the agents' states, and $n \coloneqq \sum_{i\in \mathcal V} n_i$. We call the pair $(\mathcal G, \mathbf p)$ (or equivalently,  the vector $\mathbf p$) a \textit{configuration}, and $\mathbb R^n$ is called the configuration space.

Each agent uses a suite of onboard sensors to estimate its own state and self-reports the estimated state to its neighbors. The reported state of agent $i$ is denoted by $\hat {\mathbf p}[i]$, which may or may not coincide with $\mathbf p[i]$. The discrepancy between $\mathbf p$ and $\hat {\mathbf p}$ is encapsulated in the error vector, defined as $\mathbf x \coloneqq \mathbf p - \hat {\mathbf p}$.  We say that there is a \textit{fault} or an \textit{error} at agent $i$ to mean that $\mathbf p[i] \neq \hat{\mathbf p}[i]$, or equivalently, $\mathbf x[i] \neq \mathbf 0$. This can occur if agent $i$ has wrongly estimated its state due to sensor bias, miscalibration, modeling errors, or adversarial sensor spoofing attacks.
Let $\mathcal D$ be the set of agents that have errors. We have,
\begin{equation}
|\mathcal D| = \sum_{i\in \mathcal V} \mathbb I\left(\mathbf p[i] \neq \hat{\mathbf p}[i]\right) = \|\mathbf x\|_{2,0},
\end{equation}
which is in turn
equal to the number of non-zero blocks of $\mathbf x$, referred to as its \textit{block-sparsity}.
We say that the errors are sparse, and that $\mathbf x$ is block-sparse, if 
$\|\mathbf x\|_{2,0} \ll|\mathcal V|$.

\vspace{2pt}
\begin{remark}
In the literature on rigidity theory, it is assumed that the identities of the agents in $\mathcal D^\complement$, which are called the \textit{anchors}, are known to all the agents in the network \cite{zhao2019bearing}.
However, we have dropped the assumption that any of the elements of $\mathcal D$ or $\mathcal D^\complement$ are known \textit{a priori}, allowing us to diagnose the entire network for errors, rather than ruling out a subset of the agents (i.e., the anchors) from having errors. 
\label{rem:anchors}
\end{remark}
\vspace{2pt}

In the single-agent fault detection, identification, and reconstruction (FDIR) problem, the objective is to identify the subset of sensors that are faulty, as well as recover the nominal state estimation performance. We can extend this idea to the multi-agent setting by treating each agent of the network as a \textit{sensor} that is observing the collective state vector, $\mathbf p$. With this interpretation, the onboard state estimates of the agents can also be thought of as measurements of $\mathbf p$:
\begin{align}
    \hat {\mathbf p}[i] = \bigl[\ 
        \mathbf 0 \ \ \mathbf 0 \ \ &\dots \ \ 
        \underset{
            \substack{\vspace{1pt}\\
                        \textstyle \uparrow \vspace{1pt}\\
                        \scalebox{0.8}{$i^{th}\ \textrm{block}$}
                    }
                }{\mathbf I} 
        \ \ \dots \ \ \mathbf 0\ \bigr]\ \mathbf p + \mathbf x[i], 
    \label{eq:onboard}
\end{align}
where $i\in \mathcal V$. Similarly, $\mathbf x[i]$ represents an unknown additive signal whose presence or absence must be determined at each sensor, as part of the multi-agent FDIR problem. However, the measurements in (\ref{eq:onboard}) are decoupled, as only agent $i$ is able to measure $\mathbf p[i]$ and $\mathbf x[i]$. This makes it impossible for the multi-agent network to collaboratively identify faults unless an additional set of measurements is available.

\subsection{Inter-Agent Measurements}

A distinctive feature of many real-world multi-agent systems is the availability of relative measurements between them, which are often nonlinear functions of the agents' states. We allow each hyperedge in $\mathcal E$ to correspond to a different type of nonlinear measurement.
Let the inter-agent measurement model corresponding to $\mathcal E^{(l)}$, which is the $l^{th}$ edge in $\mathcal E$, be denoted by $
\mathbf \Phi^{(l)}: U^{(l)} \rightarrow \mathbb R^{m_l}$, where $m_l$ is the dimension of the measurement and $U^{(l)}$ is an open subset of $\mathbb R^n$. Some examples of inter-agent measurement models that arise in practical applications are given in Table \ref{tab:iamms}, in each of which the state vector $\mathbf p[i]$ represents the position of agent $i$ in $2$ or $3$-dimensional space.

\begin{table}[h!]
\centering
\caption{Examples of inter-agent measurement models, where $\lbrace\mathbf p[i]\rbrace_{i\in \mathcal V}$ represent positions of the agents in $\mathbb R^2$ or $\mathbb R^3$}
\bgroup
\def\arraystretch{2.0}
\setlength{\tabcolsep}{1.0em}
{\rowcolors{2}
{gray!3}{gray!12}
\begin{tabular}{|p{2.5cm}|p{5.0cm}|}
\hline
Measurement Type & Expression for $\mathbf \Phi^{(l)}$ \\
\hline
Displacement \cite{oh2015survey} & $\ \mathbf p[i] - \mathbf p[j]$ \vspace{4pt}  \\
Distance \cite{oh2015survey,topology_const2015observability} & $\  \big\|\mathbf p[i] - \mathbf p[j]\big\|$ \vspace{4pt}  \\
Bearing \cite{zhao2019bearing} & $(\mathbf p[i] - \mathbf p[j])\Big/\big\|\mathbf p[i] - \mathbf p[j]\big\|$ \vspace{4pt}\\
Time-Difference-of-Arrival (TDoA) \cite{tdoa_2017} & $\big\|\mathbf p[i] - \mathbf p[j]\big\| 
- \big\|\mathbf p[i] - \mathbf p[k]\big\|$ \\
Subtended Angle \cite{weak_rigidity} &  $\bigstrut[t]
\arccos
\biggl(
\cfrac{\mathbf p[i] - \mathbf p[j]}{\|\mathbf p[i] - \mathbf p[j]\|}\overset{^\top}{^{\ }}
\cfrac{\mathbf p[i] - \mathbf p[k]}{\|\mathbf p[i] - \mathbf p[k]\|}
\biggr)$ \vspace{3pt} \\
\hline
\end{tabular}}
\egroup
\label{tab:iamms}
\end{table}

\begin{remark}
Other inter-agent measurement models, such as Frequency-Difference-of-Arrival (FDoA) \cite{tdoa_2017}, may involve the relative velocities between agents, in which case $\mathbf p[i]$ may be a block vector constituting the $i^{th}$ agent's position and velocity vectors. Similarly, the agents' orientations can be appended to their state vectors in order to model relative pose measurements (which can be obtained using cameras) \cite{aragues2011relative_pose}.
\end{remark}
\vspace{2pt}

For the measurement models given in Table \ref{tab:iamms}, the domain $U^{(l)}$ of the $l^{th}$ measurement model $\mathbf \Phi^{(l)}$ can be chosen to exclude the set of points in $\mathbb R^n$ where two agents' positions are coincident, as $\mathbf \Phi ^{(l)}$ and/or its derivatives may not be well-defined at these points. With the appropriate choices of the domains, each of the measurement models discussed thus far satisfy the following assumption.

\vspace{2pt}
\begin{assumption}[Smoothness]
For all $l=1, 2, \dots, |\mathcal E|$, $\mathbf \Phi^{(l)}$ is a smooth (i.e., infinitely differentiable) map.
\end{assumption}
\vspace{2pt}

The inter-agent measurements are collectively represented by the block vector-valued function,
\begin{align}
    \mathbf \Phi:\ &U \rightarrow \ \mathbb R^m\\
    & \mathbf p \mapsto \begin{bmatrix}
        \mathbf \Phi^{(1)}(\mathbf p)\\
        \mathbf \Phi^{(2)}(\mathbf p)\\
        \vdots\\
        \mathbf \Phi^{(|\mathcal E|)}(\mathbf p)
    \end{bmatrix}
\end{align}
where $U=U^{(1)}\times \dots \times U^{(|\mathcal E|)} \subseteq \mathbb R^n$ is the domain of $\mathbf \Phi$. Let $\mathbf y = \mathbf \Phi(\mathbf p)$ be the block vector of inter-agent measurements, which is partitioned such that $\mathbf y[l]= \mathbf \Phi^{(l)}(\mathbf p)$  $\forall l=1, 2, \dots, |\mathcal E|$. We assume that agent $i$ has access to the set of inter-agent measurements obtained in its neighborhood in $\mathcal G$, which is the set $\lbrace \mathbf y[l] \hsp \vert \hsp i \in \mathcal E^{(l)}\rbrace$. Letting, $\mathbf J_{\mathbf \Phi}(\mathbf p)$ denote the Jacobian of $\mathbf \Phi$ evaluated at the point $\mathbf p$, which is partitioned such that $\mathbf J_{\mathbf \Phi}(\mathbf p)[l,i]\in \mathbb R^{m_l \times n_i}$, the following assumption formalizes the functional dependence between the measurement models and the agents.

\vspace{2pt}
\begin{assumption}[Inter-Agent Sensing Topology]
We have, $\mathbf J_{\mathbf \Phi}(\mathbf p)[l, i] \neq \mathbf 0\ \Leftrightarrow\ i \in \mathcal E^{(l)}$. This means that the $l^{th}$ block in $\mathbf \Phi(\mathbf p)$ only depends on the states of the agents in $\mathcal E^{(l)}$.
\label{ass:jacobian}
\end{assumption}
\vspace{2pt}

With the above definitions in place, the objective of multi-agent FDIR can be stated as follows: the agents must collectively reconstruct an error vector $\hat{\mathbf x}\in \mathbb R^n$ that solves the equation $\mathbf y = \mathbf \Phi(\hat{\mathbf p} + \hat {\mathbf x})$, i.e., it explains the observed inter-agent measurements. Moreover, we require the error reconstruction to be done in a distributed manner, while ensuring that the computational and communication costs of the algorithm do not scale with $|\mathcal V|$. Suppose the error vector is reconstructed correctly, such that $\hat{\mathbf x} = \mathbf x$, then the indices of the non-zero blocks of $\hat{\mathbf x}$ can be identified with the faulty agents (i.e., the agents in $\mathcal D$). Finally, the true configuration of the multi-agent system can be determined as $(\mathcal G, \hat{\mathbf p} + \hat {\mathbf x})$, thereby solving the FDIR problem. However, as we show in the next section, the set of error vectors that explain the observed measurements, $\lbrace\hat{\mathbf x} \in \mathbb R^n \hsp \vert \hsp\mathbf y = \mathbf \Phi(\hat{\mathbf p} + \hat {\mathbf x})\rbrace$, may have infinitely many elements.
Therefore, additional assumptions and/or regularization techniques are needed to uniquely reconstruct ${\mathbf x}$.

%
%
\section{Error Reconstruction using Sparsity}
\label{sec:reconstruction}
Given a subset $Y\subseteq \mathbb R^m$ of the codomain of $\mathbf \Phi$, we let
 $\mathbf \Phi^{-1}\bigl[Y\bigr]$ denote its preimage under $\mathbf \Phi$, defined as follows:
 \begin{equation}
     \mathbf \Phi^{-1}\bigl[Y\bigr] = \lbrace \mathbf v \in \mathbb R^n \hsp\vert\hsp \mathbf \Phi (\mathbf v) \in Y \rbrace
 \end{equation}
Thus, $\mathbf \Phi^{-1}[\lbrace \mathbf y\rbrace]\subseteq \mathbb R^n$ is the set of all state vectors that may have generated the measurement $\mathbf y$, which we refer to as the \textit{search-space} of the error reconstruction problem. In particular, we have that $\mathbf p \in \mathbf \Phi^{-1}[\lbrace \mathbf y\rbrace]$.

For each of the inter-agent measurement models listed in Table \ref{tab:iamms}, the function $\mathbf \Phi$ is neither injective (i.e., one-one) nor surjective (i.e., onto), making it non-invertible. To see that it may fail to be injective, observe that rigid translations and rotations of Euclidean space preserve the inter-agent distances, while scaling (expansion and shrinking) of Euclidean space preserves inter-agent bearings, which means that two different configurations may generate the same measurement vector \cite{zhao2019bearing}.
To see why $\mathbf \Phi$ may fail to be surjective, consider $3$ agents organized in a triangular formation; their inter-agent distances (or displacements) must then satisfy the triangle inequality, and therefore $\mathbf \Phi$ does not attain the points in its codomain where the triangle inequality fails. Therefore, we have that $\mathbf \Phi^{-1}[\lbrace \mathbf y\rbrace] \neq \lbrace \mathbf p \rbrace$ in general. Nevertheless, it is still possible to uniquely reconstruct $\mathbf p$ using $\mathbf y$ if additional assumptions are introduced. 

In the remainder of this section, we use tools from differential geometry to show that $\mathbf x$ may be uniquely reconstructed if the errors are assumed to be sparse, i.e., $|\mathcal D|\ll|\mathcal V|$. We also draw parallels between the main result of this section (Theorem \ref{thm:generalized}) and the existing literature on rigidity theory, showing that our analysis generalizes the concept of rigidity in distance and bearing-based multi-agent localization (which are both special cases of our problem formulation).
\subsection{Characterization of the Search-Space}
\label{subsec:search-space}
Following the definitions in \cite[p. 182]{lee2012}, a vector $\tilde {\mathbf p} \in \mathbb R^n$ is said to be a \textit{regular point} of $\mathbf \Phi$ if $\mathbf J_{\mathbf \Phi}(\tilde {\mathbf p})$ has full row rank, i.e., is a surjective map. A \textit{regular value} of $\mathbf \Phi$ is a vector $\tilde {\mathbf y}\in \mathbb R^m$ in the codomain, such that every element of $\mathbf \Phi^{-1}[\lbrace \tilde{\mathbf y} \rbrace]$ is a regular point. The relevance of regular values is due to the following lemma.


\vspace{2pt}
\begin{lemma}[Regular Level Set Theorem \protect{\cite[Corr. 8.10]{lee2012}}]
If $\tilde {\mathbf y}$ is a regular value of $\mathbf \Phi$, then $\mathbf \Phi^{-1}[\lbrace \tilde{\mathbf y} \rbrace]$ is an $(n-m)$-dimensional embedded submanifold of $\mathbb R^n$.
\label{lem:regular}
\end{lemma}
\vspace{2pt}

Lemma \ref{lem:regular} can be used to show that, while the measurement vector $\mathbf y$ does not uniquely determine the configuration, it effectively restricts the set of possible configurations (which is given by the preimage, $\mathbf \Phi^{-1}[\lbrace {\mathbf y}\rbrace]$), to an embedded submanifold in $\mathbb R^n$, i.e., a smooth `surface'.
This means that we can reconstruct the true configuration $(\mathcal G, \mathbf p)$ more efficiently by restricting our search-space to this lower-dimensional submanifold, by starting at a point on this submanifold and moving in tangential directions to search for the true configuration.

\vspace{2pt}
\begin{theorem}[Characterization of the Search-Space]
Let $k$ denote the rank of $\mathbf J_{\mathbf \Phi}(\mathbf p)$. There exists a point $\tilde{\mathbf p}$ arbitrarily close to $\mathbf p$,  such that the points in $\mathbf \Phi^{-1}[\lbrace   {\mathbf \Phi(\tilde{\mathbf p})}\rbrace]$ lie on an $(n-k)$-dimensional embedded submanifold of $\mathbb R^n$.
\label{thm:generalized}
\end{theorem}
\begin{proof}
As $\mathbf J_{\mathbf \Phi}(\mathbf p)$ is a rank $k$ matrix, we can construct a subset $\mathcal I \subseteq \lbrace 1, 2, \dots, m\rbrace$ consisting of $k$ elements, such that the rows of $\mathbf J_{\mathbf \Phi}({\mathbf p})$ which are indexed by $\mathcal I$ are linearly independent. Thereafter, define the vector-valued map $\mathbf \Phi_{\mathcal I}:\mathbb R^n \rightarrow \mathbb R^k$ whose components are precisely the components of $\mathbf \Phi(\cdot)$ which are indexed by $\mathcal I$. Let $\mathbf y_{\mathcal I}\in \mathbb R^k$ denote the vector constructed by choosing the components of $\mathbf y$ which are indexed by $\mathcal I$.

We observe that ${\mathbf p}$ is a regular point of $\mathbf \Phi_{\mathcal I}$. However, the application of Lemma \ref{lem:regular} requires ${\mathbf y}_{\mathcal I}$ to be a regular value, which may not be the case.
To fix this issue, we invoke \textit{Sard's theorem} \cite[Thm. 10.7]{lee2012}, which states that the set of elements of $\mathbb R^k$ that are not regular values of $\mathbf \Phi_{\mathcal I}  $ constitute a measure-zero subset of $\mathbb R^k$. Thus, any (arbitrarily small) open neighborhood of ${\mathbf y}_{\mathcal I}$ in $\mathbb R^k$ contains a regular value in it; in fact, almost all vectors in such a neighborhood are regular values. As $\mathbf \Phi_{\mathcal I}$ is surjective in the neighborhood of $\mathbf p$ \cite[Prop. 2]{Asimow1978}, there exists a point $\tilde {\mathbf p}$ arbitrarily close to $\mathbf p$ that maps to a regular value, i.e., $\mathbf \Phi_{\mathcal I}(\tilde{\mathbf p})$ is a regular value of $\mathbf \Phi_{\mathcal I}$. Finally, observe that
\begin{equation}
\mathbf \Phi^{-1} \bigl[\lbrace  \mathbf \Phi(\tilde{\mathbf p}) \rbrace\bigr] \subseteq \mathbf \Phi_{\mathcal I}^{-1} \bigl[\lbrace  \mathbf \Phi_{\mathcal I}(\tilde{\mathbf p})\rbrace\bigr].
\label{eq:subsets}
\end{equation}
Applying Lemma \ref{lem:regular} to the right-hand side of (\ref{eq:subsets})
completes the proof.
\end{proof}
\vspace{2pt}

\vspace{2pt}
\begin{remark}[Regular Values are Almost-Everywhere]
When $\mathbf y$ has some amount of randomness in each of its components (e.g., due to measurement noise), Sard's theorem establishes that the vector $\mathbf y_{\mathcal I}$ is a regular value of $\mathbf \Phi_{\mathcal I}$ with probability one.
Hence, the fact that $\tilde{\mathbf p}$ is not identically equal to $\mathbf p$ in the claim of Theorem \ref{thm:generalized} can be viewed as a technicality. Similar technicalities arise in the study of multi-agent localization, which motivates the study of \textit{generic} configurations, as opposed to the special, measure-zero subset of configurations for which certain properties may fail to hold \cite{whiteley1985generating, anderson2010formal, hendrickson1992}. 
\label{rem:technicality}
\end{remark}
\subsection{Connections to Rigidity Theory}
\label{subsec:example}
A special case of the error reconstruction problem is that of localizing a multi-agent system using inter-agent measurements, which has been previously studied using \textit{rigidity theory} \cite{zhao2019bearing}. In this subsection, we consider the specific example of multi-agent localization using inter-agent distances, using it to illustrate the relevance of Theorem \ref{thm:generalized}, as well as motivate our forthcoming analysis. 

Consider the example where $\mathbf p[i]\in \mathbb R^3$ is the $i^{th}$ agent's position vector and each component of $\mathbf \Phi$ is of the form
\begin{equation}
    \mathbf \Phi^{(l)}(\mathbf p)=\|\mathbf p[i] - \mathbf p[j]\|
\end{equation}
where $\mathcal E^{(l)}=\lbrace i, j\rbrace$. This gives us the dimensions $n=3|\mathcal V|$ and $m=|\mathcal E|$ of the domain and codomain of $\mathbf \Phi$, respectively. In this case, the matrix $\mathbf J_{\mathbf \Phi}(\mathbf p)\in \mathbb R^{|\mathcal E|\times 3|\mathcal V|}$ is called the \textit{rigidity matrix}, and its maximal rank is well-known to be equal to $(3|\mathcal V|-6)$ \cite{eren2004rigidity, zelazo2015decent, hendrickson1992}. A multi-agent configuration is said to be \textit{infinitesimally rigid} if the rank of its rigidity matrix is maximal. Similar to our observation in Remark \ref{rem:technicality}, infinitesimal rigidity fails to hold for a measure-zero subset of the configuration space \cite[Thm. 2.1]{hendrickson1992}. 

Suppose an infinitesimally rigid configuration generates the inter-agent distance vector, $\mathbf y \in \mathbb R^{|\mathcal E|}$. Using Theorem \ref{thm:generalized} with $n=3|\mathcal V|$ and $k=(3|\mathcal V|-6)$, we see that the set of all configurations that generate the measurement $\mathbf y$ constitute a $6$-dimensional embedded submanifold of $\mathbb R^3$. Denote this submanifold as $\mathcal M_{\mathbf y}$. 
As rigid translations and rotations (i.e., the isometries) of the multi-agent configuration preserve the inter-agent distances, it follows that each connected component of $\mathcal M_{\mathbf y}$ corresponds to the $SE(3)$ Lie group\footnote{To state this more precisely, each of the connected components of $\mathcal M_{\mathbf y}$ is the orbit of some configuration $(\mathcal G,\tilde{\mathbf p})$ under $SE(3)$ equipped with an appropriate Lie group action on $\mathbb R^n$. A similar observation is made in \cite[Lemma 3]{krick2009stabilisation} for the two-dimensional case.} (which are precisely the isometries of the three-dimensional Euclidean space, excluding reflections). Indeed, $SE(3)$ is 6-dimensional as a manifold \cite[Sec. 1.2.5]{hall2013lie}. 

\begin{figure}
     \centering
     \begin{minipage}[b]{0.22\textwidth}
         \centering
         \includegraphics[trim={0.95cm 1.0cm 0.95cm 1.0cm},clip, width=\textwidth]{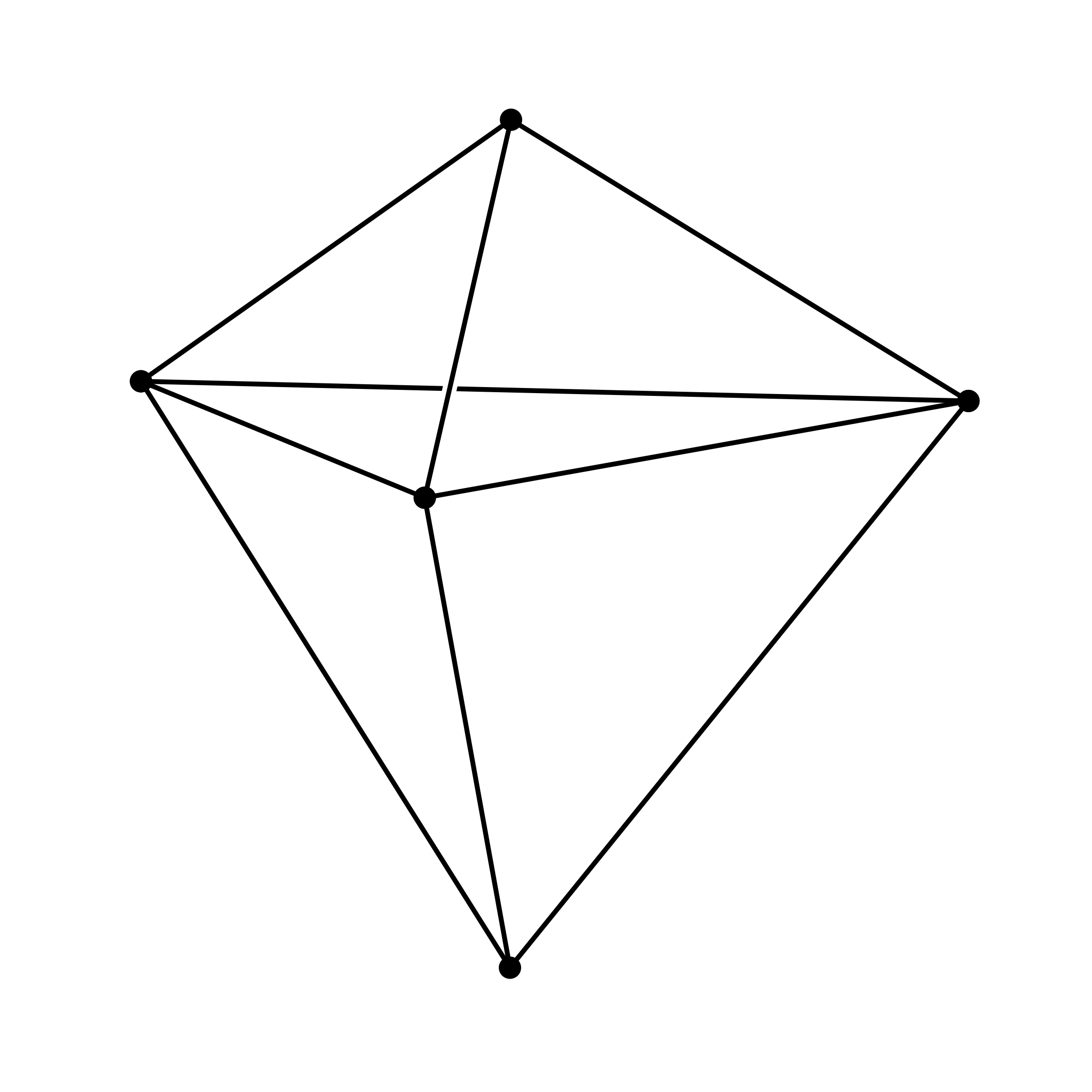}
     \end{minipage}
     \ 
     \begin{minipage}[b]{0.22\textwidth}
         \centering
         \includegraphics[trim={0.95cm 1.0cm 0.95cm 1.0cm},clip, width=\textwidth]{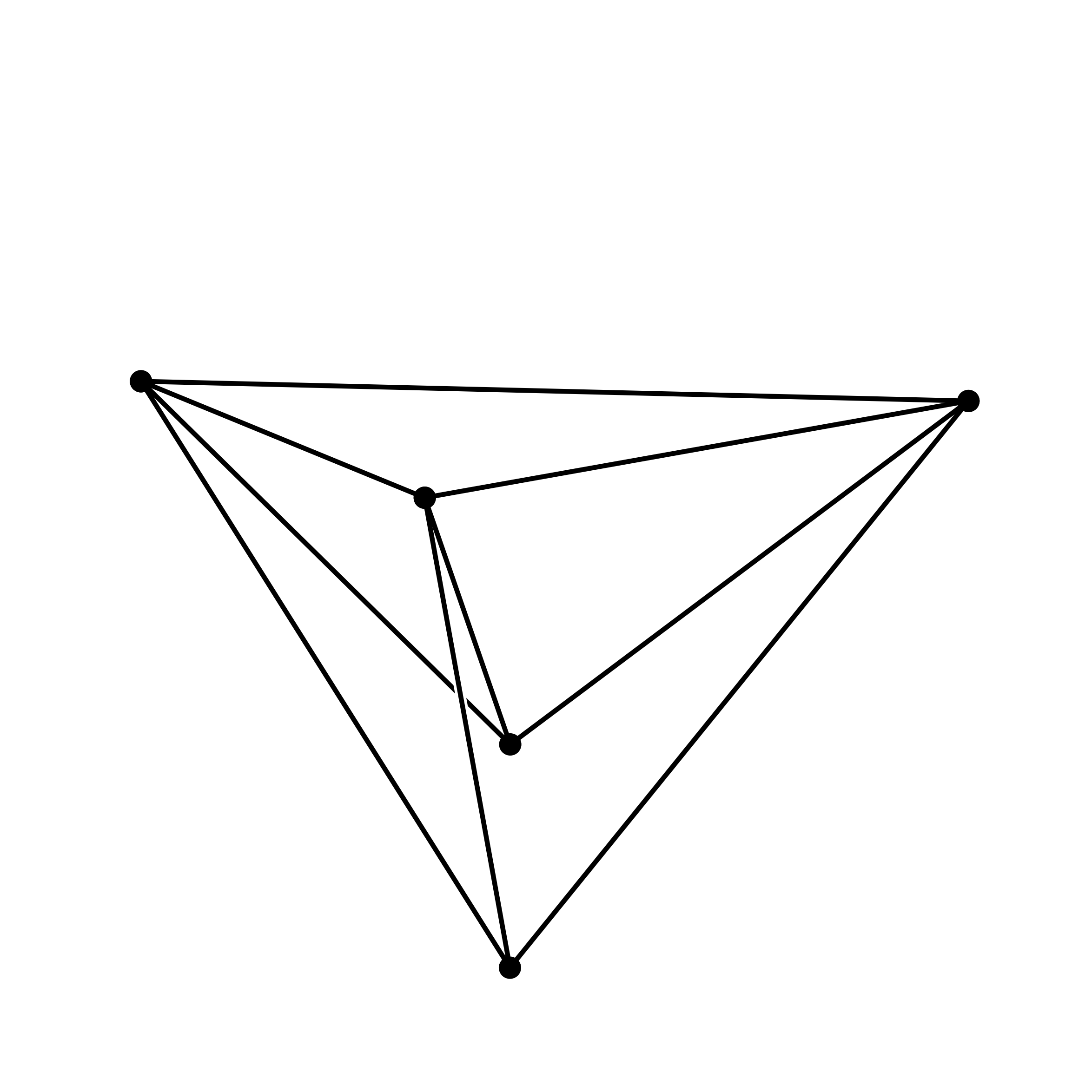}
     \end{minipage}
        \caption{Two configurations of a rigid graph that produce the same measurement vector $\mathbf y$, but are not related to each other via rigid translations or rotations, are said to be \textit{flip ambiguous}.}
        \label{fig:example}
\end{figure}

Note that the manifold $\mathcal M_{\mathbf y}$ may have two or more disconnected components. The disconnected components of $\mathcal M_{\mathbf y}$ correspond to what are called \textit{flip ambiguities} in the literature on rigidity theory, which is the phenomenon that is visualized in Fig. \ref{fig:example}. The two configurations shown in Fig. \ref{fig:example} both lie on $\mathcal M_{\mathbf y}$ as they produce the same vector of inter-agent distances, $\mathbf y$. However, there is no continuous path on $\mathcal M_{\mathbf y}$ that connects the two configurations in $\mathbb R^{3|\mathcal V|}$, for the same reason that no continuous motion of the vertices of one configuration can make it into the other while preserving the inter-agent distance measurements.

As we mentioned in Remark \ref{rem:anchors}, the existing literature assumes that the anchors are known to be correctly localized. Equivalently, they assume that the components of $\mathbf x$ which correspond to the anchors are equal to $\mathbf 0$. Observe that the foregoing assumption defines a subspace of the configuration space. By intersecting this subspace with $\mathcal M_{\mathbf y}$, the true configuration may be recovered (up to flip ambiguity), as the intersection of two manifolds is (typically) a lower-dimensional manifold \cite[p. 203]{lee2012}. In particular, a zero-dimensional manifold is a set of discrete points that correspond to configurations that are infinitesimally rigid.

In the following subsection, we show that the assumption that the anchors' identities are known (i.e., that the locations of the non-zero blocks of $\mathbf x$ are known) can be relaxed in favor of a milder assumption: the errors are sparse (i.e., some of the blocks of $\mathbf x$ are equal to $\mathbf 0$). The latter assumption is less restrictive in the context of the FDIR problem, as none of the agents' states need to be trusted, but rather, the entire multi-agent network may be diagnosed for faults.

\subsection{The Role of Sparsity}
\label{subsec:sparsity}
Let $\mathcal M_{\mathbf y}$ be the submanifold of $\mathbb R^n$ which contains the true state vector $\mathbf p$ on it. Denote by $(\mathcal M_{\mathbf y} - \hat{\mathbf p})$ the manifold constructed by translating each point of $\mathcal M_{\mathbf y}$ by $(-\hat{\mathbf p})$; equivalently, we may conceptualize it as the shifting of the origin of $\mathbb R^n$ to $\hat{\mathbf p}$. Observe that $(\mathcal M_{\mathbf y} - \hat{\mathbf p})$ is the manifold of all the error vectors which can explain the observed measurements, $\mathbf y$. In particular, we have $\mathbf x \in (\mathcal M_{\mathbf y} - \hat{\mathbf p})$.

\begin{figure}
         \centering
         \includegraphics[width=0.30\textwidth]{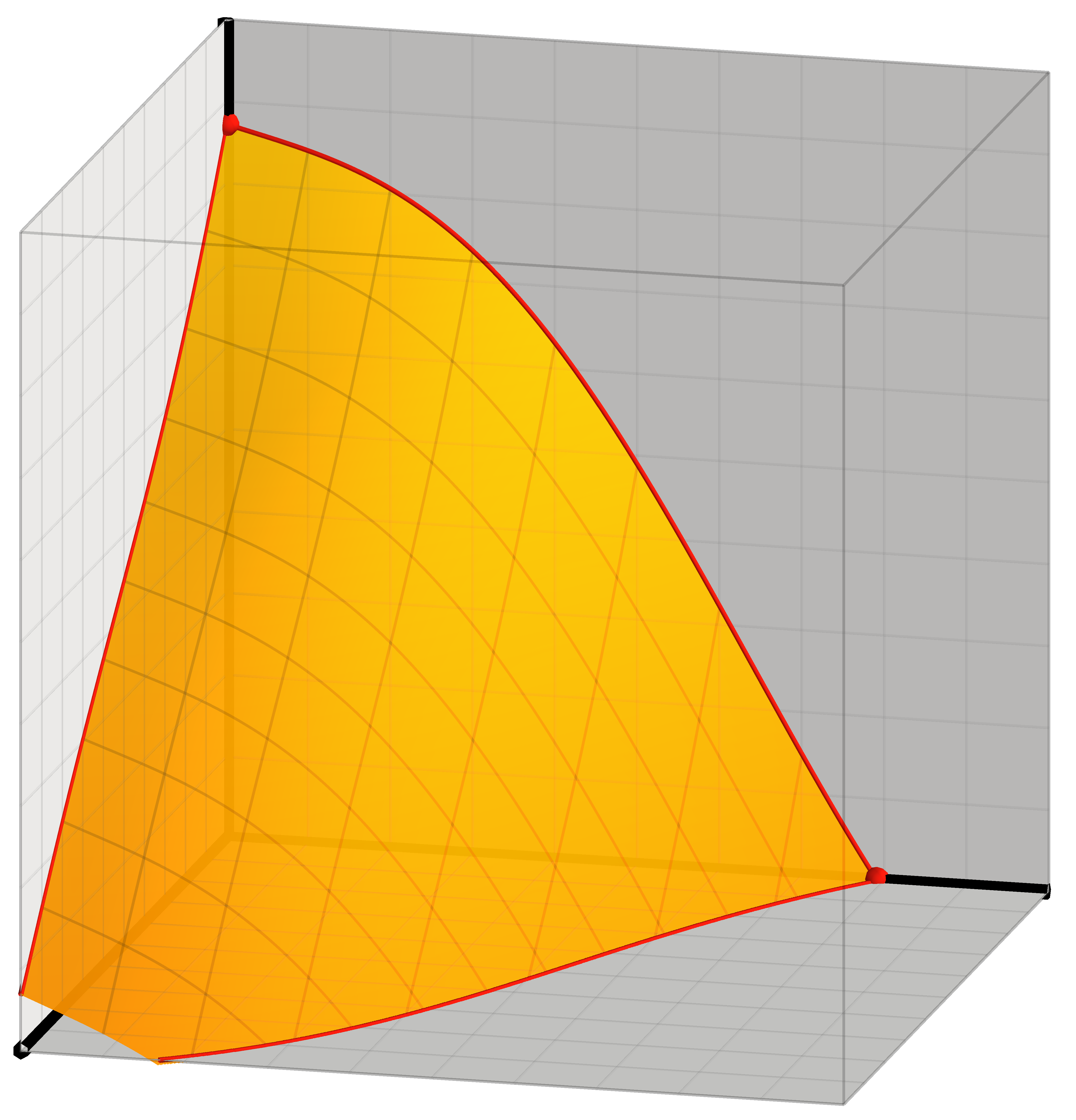}
        \caption{The intersection of a generic two-dimensional manifold (yellow surface) with the set of $2$-sparse vectors is a one-dimensional manifold (red curves). Its intersection with the set of $1$-sparse vectors is a zero-dimensional manifold (red points).}
        \label{fig:sparsity}
\end{figure}

Given a nonnegative integer $s< n$, consider the set of $s$-sparse vectors in $\mathbb R^n$, by which we mean the set of vectors that have at most $s$ non-zero elements. The set of $1$-sparse vectors are precisely the axes of $\mathbb R^n$, i.e., the union of each of the $1$-dimensional subspaces spanned by the standard basis vectors of $\mathbb R^n$. More generally, the set of $s$-sparse vectors is a union of $s$-dimensional subspaces \cite[p. 20]{wakin2007geometry_p20}. By a similar argument, the set of $s$ block-sparse vectors of $\mathbb R^n$ (whose block-sparsity is at most $s$, where $s<|\mathcal V|$) is a union of lower-dimensional subspaces. Therefore, if the errors are known to be sparse, we can intersect the set of $s$ block-sparse vectors with the manifold $(\mathcal M_{\mathbf y} - \hat{\mathbf p})$, significantly reducing our search space; this idea is illustrated in Fig. \ref{fig:sparsity}. However, since we do not assume that the set $\mathcal D$ or its cardinality is known, the block-sparsity of $\mathbf x$ is unknown as well.

The discussion thus far motivates the following optimization problem for simultaneously identifying the faulty agents as well as recovering their corresponding error vectors:
\begin{align*}
\begin{array}{rclc}
   \textsc{P1:}  &  &
\begin{array}{rl}
\underset{
\scalebox{0.9}{$\hat{\mathbf x}\hsp\in\hsp\mathbb R^{n}$}}
{\textnormal{minimize\ }}\quad  
&\| \hat{\mathbf x}\|_{2,0}  
\\
\textnormal{\bigstrut[t] subject to\ } \quad
& \mathbf \Phi \left(\hat {\mathbf p} + \hat{\mathbf x}\right) \hsp = \hsp  \mathbf y 
\end{array} & \qquad \qquad
\end{array}
\end{align*}
Problem P1 identifies the minimum number of faults that are able to explain the observed measurement vector $\mathbf y$. This is a natural way to pose the FDIR problem when the occurrence of faults at each of the agents are independent and identically distributed (i.i.d.) processes, such that given a nonnegative integer $s<|\mathcal V|$, the likelihood of the occurrence of $s+1$ faults is strictly smaller than that of $s$ faults. Nevertheless, our present motivation for considering Problem P1 is that it has the potential\footnote{The condition for the intersection of two submanifolds of $\mathbb R^n$ to be a lower-dimensional submanifold is called \textit{transversality}; it can be found in \cite[p. 68]{bott1982differential}.} to reduce the dimension of the search space. In special cases (e.g., the distance-based localization problem considered in Section \ref{subsec:example}), it is possible to determine the exact conditions under which an $s$ block-sparse error can be uniquely recovered in this manner \cite{khan2023recovery}.


%
%
\section{Distributed FDIR using ADMM}
\label{sec:distributed}
As Problem P1 is combinatorial in its nature, it invariably requires a series of reformulations before it may be solved efficiently. Moreover, while the objective function of P1 is separable, as $\|\mathbf x\|_{2,0}=\sum_{i\in\mathcal V}\|\mathbf x[i]\|_{2,0}$, the equality constraint of P1 introduces a coupling between the blocks of $\mathbf x$,
preventing P1 from being separable between the agents in $\mathcal V$. In \cite{blumensath2013compressed}, an algorithm based on Iterative Hard Thresholding (IHT) was proposed for solving P1; however, it requires a high-dimensional projection operation that is infeasible for large-scale multi-agent applications.

In this section, we develop a distributed algorithm for solving P1 by using a combination of the sequential convex programming (SCP) and alternating direction method of multipliers (ADMM) techniques. 
Additionally, we exploit the structure of the graph (which is encapsulated by Assumption \ref{ass:jacobian}) to ensure that computational and communication costs of the resulting algorithm do not scale with the size of the graph.
\subsection{Reduction to Convex Sub-Problems}
\label{subsec:scp}
The objective of SCP is to solve a nonconvex optimization problem by breaking it down into convex sub-problems, motivated by the ease and speed of solving convex optimization problems \cite{scp_zillober2004,scp_morgan2014}.  To arrive at the convex sub-problem, we first linearize the nonlinear constraint of P1 as follows.
Let $\mathbf x^*\in\mathbb R^n$ denote a partially reconstructed error vector which has been computed over a given number of SCP iterations. The subsequent iteration of SCP searches for a block vector $\hat{\mathbf x}\in \mathbb R^n$ satisfying $\mathbf \Phi( \hat{\mathbf p} + \mathbf x^* + \hat {\mathbf x})=\mathbf y$, where $(\hat{\mathbf p} + \mathbf x^*)$ is treated as an updated estimate of $\mathbf p$.
The first-order Taylor series approximation of $\mathbf \Phi$ near the point $(\hat{\mathbf p} + \mathbf x^*)$ is given by
\begin{align}
\mathbf \Phi( \hat{\mathbf p} + \mathbf x^* + \hat {\mathbf x}) \approx \mathbf \Phi(\hat{\mathbf p} + \mathbf x^*)  + \mathbf J_{\mathbf \Phi} (\hat{\mathbf p} + \mathbf x^*) (\hat {\mathbf x})
\label{eq:linearization}
\end{align}
Defining $\mathbf z \coloneqq \mathbf y - \mathbf \Phi(\hat{\mathbf p} + \mathbf x^*)$ and $\mathbf R \coloneqq \mathbf J_{\mathbf \Phi}(\hat{\mathbf p} + \mathbf x^*)$, we arrive at a linearized version of the equality constraint, $\mathbf R \hat{\mathbf x} = \mathbf z$.

The objective function of P1, $\|\hsp\cdot\hsp\|_{2,0}$, is still a non-convex function. Thus, consider replacing it with the convex function, $\|\hsp\cdot\hsp\|_{2,1}$. The justification for this is as follows; when the matrix $\mathbf R$ satisfies a certain numerical property, called the \textit{block null space property}, the objective functions $\| \hsp \cdot \hsp \|_{2,0}$ and $\|\hsp \cdot \hsp \|_{2,1}$ both yield the same minimum \cite[Thm. 1]{robust_NSP_2017}. 

\vspace{2pt}
\begin{remark}
The exact conditions on the multi-agent configuration under which the block null space property holds can be found in \cite{khan2023recovery}, where it was studied for the special case of multi-agent localization using distance or bearing measurements. 
As for the general FDIR problem being considered in this paper, our analysis in Section \ref{subsec:interpretation} reveals that $\|\hsp \cdot \hsp \|_{2,1}$ promotes the block sparsity of the solution irrespective of whether any conditions on $\mathbf R$ are met.
\end{remark}
\vspace{2pt}

Together, the two preceding steps lead us to the following optimization problem:
\begin{align*}
\begin{array}{rclc}
   \textsc{P2:}  &  &
\begin{array}{rl}
\underset{
\scalebox{0.9}{$\hat{\mathbf x}\hsp\in\hsp\mathbb R^{n}$}}
{\textnormal{minimize\ }}\quad  
&\| \mathbf x^* + \hat{\mathbf x}\|_{2,1}  
\\
\textnormal{\bigstrut[t] subject to\ } \quad
& \mathbf R \hat{\mathbf x} = \mathbf z
\end{array} & \qquad \qquad
\end{array}
\end{align*}
which is indeed convex. 
In each iteration of SCP, one first solves the linearized sub-problem P2, while keeping $\mathbf x^*$, $\mathbf z$, and $\mathbf R$ fixed, and then updates $\mathbf x^*$ by adding to it the minimizer of P2. In the next SCP iteration, the constraint is re-linearized using the new value of $\mathbf x^*$.
By repeatedly performing these two steps (i.e., minimization followed by re-linearization), a locally optimal solution to the original nonconvex optimization problem can be obtained \cite[Prop. 2]{khan2023recovery}.

\vspace{2pt}
\begin{remark}
The sub-problem P2 may be infeasible due to the linearization error in the constraint. To resolve this, slackness can be introduced in the constraint \cite{khan2023recovery} or a trust-region method can be used \cite[Sec. 6]{bonalli2022sequential}. In our numerical examples in Sec. \ref{sec:numerical}, it is seen that approximate minimization of P2 also works quite well in practice. In fact, ADMM is particularly well-suited for fast approximate minimization of a convex optimization problem \cite[Sec. 3.2.2]{boyd2011distributed}, which motivates the use of ADMM to solve the sub-problem P2.
\end{remark}
\vspace{2pt}

However, the equality constraint of P2 still poses a challenge, as it
causes the optimization problem to be coupled between the agents. In the next subsection, we further separate Problem P2 into sub-problems that can be solved in parallel by the agents in the network. 
\subsection{Derivation of the Distributed Algorithm}
\label{subsec:admm}
ADMM uses the theory of Lagrangian duality to solve convex optimization problems. Specifically, it introduces another set of primal variables (in addition to $\hat {\mathbf x}$) to decouple the optimization problem, and then augments the Lagrangian of the problem with a quadratic penalty term. To this end, we introduce the following set of primal variables corresponding to the $i^{th}$ agent: $\lbrace \hat{\mathbf w}_i^{(j)}\in \mathbb R^{n_i} \hsp |\hspace{2pt} j \in \mathcal N_i\rbrace$, whose elements are meant to represent $|\mathcal N_i|$ duplicate copies of agent $i$'s error vector. Along with the new set of primal variables, we introduce an equal number of \textit{consistency constraints}:
\begin{align}
   \hat{\mathbf w}_i^{(j)} = \hat{\mathbf x} [i], \quad \forall j \in \mathcal N_i
   \label{eq:consistency_constraints}
\end{align}
Define $\hat {\mathbf w}\coloneqq \lbrace \hat{\mathbf w}_i^{(j)} \hsp | \hsp i\in \mathcal V, j\in\mathcal N_i \rbrace$ as the set of all the newly introduced primal variables, and let $\mathcal E_i\coloneqq \lbrace l \hsp | \hsp i\in\mathcal E^{(l)}\rbrace$ be the set of indices of the hyperedges that contain agent $i$.
By introducing the new primal variables and constraints into Problem P2 and then augmenting the objective function with quadratic penalty terms, we arrive at the following optimization problem:
\vspace{-35pt}
\begin{align*}
\begin{array}{rl}
    \textsc{P3:}   &
    \begin{array}{rl}
        \underset{
        \scalebox{0.9}{$\hat{\mathbf x}, \hat{\mathbf w}$}
        }{\textnormal{minimize\ }}  
        & 
        \begin{array}{l}
            \\\vspace{7pt}\\
            \hspace{-5pt}
            \|\mathbf x^* + \hat{\mathbf x}\|_{2,1} 
            + \frac{\rho}{2}\sum_{
                \begin{subarray}{l}
                l\in\mathcal E_i\\ i\in\mathcal V \end{subarray}
            }\|\mathbf c_i^{(l)}(\hat{\mathbf x}, \hat{\mathbf w})\|^2 \bigstrut[b] \vspace{4pt}\\ 
            \hspace{24.25pt} + \bigstrut[t]\frac{\rho}{2}\sum_{
                \begin{subarray}{l}
                j\in\mathcal N_i\\ i\in\mathcal V \end{subarray}
            }
            \|\mathbf d_i^{(j)}(\hat{\mathbf x}, \hat{\mathbf w})\|^2
            \vspace{8pt}
        \end{array}
        \\ 
        \textnormal{\bigstrut[t] subject to
        } \quad
        & \big\lbrace \mathbf c_i^{(l)}(\hat{\mathbf x}, \hat{\mathbf w})= \mathbf 0 \big\rbrace_{
        \begin{subarray}{l}
                l\in\mathcal E_i\\ i\in\mathcal V \end{subarray}} 
        \vspace{4pt}
        \\& 
        \big\lbrace \mathbf d_i^{(j)}(\hat{\mathbf x}, \hat{\mathbf w})=\mathbf 0\big\rbrace_{
        \begin{subarray}{l}
                j\in\mathcal N_i\\ i\in\mathcal V \end{subarray}}
    \end{array} 
\end{array}
\end{align*}
where $\rho\in\mathbb R$ is a positive real number called the penalty parameter, and 
\begin{align}
\mathbf c_i^{(l)}(\hat{\mathbf x}, \hat{\mathbf w}) &\coloneqq \mathbf R[l, i]\hat {\mathbf x}[i]-\Big(\mathbf z[l] - \sum_{j\in\mathcal N_i}\mathbf R[l,j] \hat{\mathbf w}_{j}^{(i)}\Big) 
\nonumber\\
\mathbf d_i^{(j)}(\hat{\mathbf x}, \hat{\mathbf w}) &\coloneqq \hat{\mathbf x}[i] - \hat{\mathbf w}_i^{(j)}
\label{eq:d_ij}
\end{align}
are used to represent the constraints.
The first set of constraints in P3 is a decomposed version of the constraint $\mathbf R \hat {\mathbf x} = \mathbf z$ of P2. In fact, we have the following equivalence between Problems P2 and P3.

\begin{proposition}
    Suppose $(\bar{\mathbf x}, \bar{\mathbf w})$ is a minimizer of Problem P3, then $\bar{\mathbf x}$ also minimizes Problem P2.
\end{proposition}
\begin{proof}
As a minimizer $(\bar{\mathbf x}, \bar{\mathbf w})$ of P3 must by definition be feasible, the last two terms of the objective function of P3 vanish. Since $\mathbf d_i^{(j)}(\bar{\mathbf x}, \bar {\mathbf w})=\mathbf 0$, we have that $ \bar{\mathbf w}_i^{(j)}=\bar{\mathbf x}[i]$, for all $i$ and $j$.
    Substituting this in the other constraint equation, we have 
    \begin{align}
    \mathbf c_i^{(l)}(\bar{\mathbf x}, \bar{\mathbf w})&=\sum_{j\in\mathcal N_i \cup \lbrace i\rbrace}\mathbf R[l, j]\bar{\mathbf x}[j] - \mathbf z[l]\\ 
    &= \quad \sum_{j\in\mathcal V} \quad\mathbf R[l, j]\bar{\mathbf x}[j] - \mathbf z[l] = \mathbf 0
    \label{eq:prop_used_assumption}
    \end{align}
In (\ref{eq:prop_used_assumption}), we extended the summation to include terms which are identically $\mathbf 0$, due to the corresponding blocks of $\mathbf R$ being $\mathbf 0$ (cf. Assumption \ref{ass:jacobian}). Thus, it can be seen that
\begin{align}
\big\lbrace \mathbf c_i^{(l)}(\bar{\mathbf x}, \bar{\mathbf w})= \mathbf 0 \big\rbrace_{
        \begin{subarray}{l}
                l\in\mathcal E_i\\ i\in\mathcal V \end{subarray}}
\ \Rightarrow \ \mathbf R \bar{\mathbf x} = \mathbf z
\end{align}
\end{proof}


Denote the Lagrangian of P3 by $L(\hat {\mathbf x}, \hat{\mathbf w}, \boldsymbol \lambda, \boldsymbol \mu)$, where
\begin{align}
{\boldsymbol \lambda} &\coloneqq \left\lbrace \hsp
{\boldsymbol \lambda}_{i}^{(l)} \in \mathbb R^{m_l} \hsp\big|\hspace{3pt}  
i \in \mathcal V,\hspace{3pt}
l \in \lbrace 1, \dots, |\mathcal E|\rbrace,\hspace{3pt} i\in \mathcal E^{(l)}
\right\rbrace\nonumber \\
{\boldsymbol \mu} &\coloneqq \left\lbrace \hsp
{\boldsymbol \mu}_{i}^{(j)} \in \mathbb R^{n_i}\hsp\big|\hspace{3pt}  
i \in \mathcal V,\hspace{3pt}
j\in \mathcal N_i
\right\rbrace
\end{align}
are the sets of dual variables (or Lagrange multipliers) corresponding to the two sets of constraints in P3. In each iteration of ADMM, the Lagrangian $L(\hat {\mathbf x}, \hat{\mathbf w}, \boldsymbol \lambda, \boldsymbol \mu)$ is first minimized with respect to $\hat{\mathbf x}$, then it is minimized with respect to $\hat{\mathbf w}$, 
and finally, a gradient ascent step is used to update the dual variables \cite{boyd2011distributed}. 
Our next task of order is to show that each of the primal minimization steps can be carried out in a distributed manner. In the following, we omit the arguments of $L$, ${\mathbf c_i^{(l)}}$ and ${\mathbf d_i^{(j)}}$ for brevity.
\begin{proposition}
The Lagrangian of P3 can be expressed in the following, equivalent forms:
\begin{align}
L &=
\sum_{i\in\mathcal V}
\bigg[
\|\hat{\mathbf x}[i]+\mathbf x^*[i]\| 
+\sum_{l\in\mathcal E_i} \left(
\frac{\rho}{2}\|\mathbf c_i^{(l)}\|^2 + \boldsymbol{\lambda}_i^{(l)
\raisebox{2pt}{$\scriptstyle\top$}
}
\mathbf c_i^{(l)} \right)
\nonumber\\
&\hspace{75pt} +
\sum_{j\in\mathcal N_i} \left( \frac{\rho}{2}\|
\mathbf d_i^{(j)}
\|^2 + \boldsymbol{\mu}_i^{(j)
\raisebox{2pt}{$\scriptstyle\top$}
}
\mathbf d_i^{(j)}\right)\bigg]
\label{eq:L'_i_def}\\
&= \sum_{i\in\mathcal V}\bigg[
\ \cdots \ +
\sum_{j\in\mathcal N_i} \left( \frac{\rho}{2}\|
\mathbf d_j^{(i)}
\|^2 + \boldsymbol{\mu}_j^{(i)
\raisebox{2pt}{$\scriptstyle\top$}
}
\mathbf d_j^{(i)}\right) \bigg]
\label{eq:L'_i_def_2}
\end{align}
where `$\cdots\hspace{0.5pt}$' is used to indicate that the first two terms inside the summation $\sum_{i\in\mathcal V}(\hsp\cdot\hsp)$ are identical in both expressions.
\end{proposition}
\begin{proof}
Firstly, note that $\|\mathbf x\|_{2,1}=\sum_{i\in\mathcal V}\|\mathbf x[i]\|$, which follows from the definition of $\|\hsp\cdot\hsp\|_{2,1}$. Thus, each term of the objective function of P3 has a summation of the form $\sum_{i\in\mathcal V}(\hsp\cdot\hsp)$ around it. It is also clear by comparing P3 and (\ref{eq:L'_i_def}) how the constraints are being split between the agents, as well as the pairing between the constraints and the Lagrange multipliers. It only remains to be seen that the Lagrangian can be equivalently expressed in the two different forms that are given in (\ref{eq:L'_i_def}) and (\ref{eq:L'_i_def_2}). To see this, observe that
\begin{align}
   \sum_{i\in\mathcal V} \sum_{j\in\mathcal N_i} &\Big( \frac{\rho}{2}\|
\mathbf d_i^{(j)}
\|^2 + \boldsymbol{\mu}_i^{(j)
\raisebox{2pt}{$\scriptstyle\top$}
}
\mathbf d_i^{(j)}\Big)\nonumber\\
&=\sum_{j\in\mathcal V} \sum_{i\in\mathcal N_j} \Big( \frac{\rho}{2}\|
\mathbf d_i^{(j)}
\|^2 + \boldsymbol{\mu}_i^{(j)
\raisebox{2pt}{$\scriptstyle\top$}
}
\mathbf d_i^{(j)}\Big)
\label{eq:L'_equivalence}
\end{align}
which is true because the summations $\sum_{i\in\mathcal V}\sum_{j\in\mathcal N_i}(\hsp \cdot \hsp)$ and  $\sum_{j\in\mathcal V}\sum_{i\in\mathcal N_j}(\hsp \cdot \hsp)$ both sum over the same set of terms.
By interchanging the indices on the right-hand side of (\ref{eq:L'_equivalence}), we see that the two decompositions of $L$ are equivalent.
\end{proof}

By decomposing the Lagrangian appropriately, the agents can execute the primal minimization steps in parallel, without needing to perform any redundant computations (e.g., due to two or more agents minimizing over the same variable).
The overall distributed multi-agent FDIR protocol which combines the SCP and ADMM optimization techniques is presented in Algorithm \ref{alg:admm}. The proposed algorithm has two loops: the outer loop corresponds to SCP, and the inner loop corresponds to the ADMM subroutine (which is presented separately, in Alg. \ref{alg:inner}). 
To implement the proposed FDIR algorithm, all the agents must know the penalty parameter $\rho$ and the inter-agent communications must be synchronous; these requirements can be fulfilled by using the distributed algorithms given in
\cite{olshevsky2009convergence} and \cite{sync2010}, respectively.



In either of the primal minimization steps of the proposed FDIR algorithm at agent $i$ (i.e., steps 3 and 5 of Alg. \ref{alg:inner}), the agent solves a convex optimization problem whose dimension scales linearly with the size of the agent's neighborhood in $\mathcal G$, i.e., $|\mathcal N_i|$, which can be considered as a measure of how densely connected the network is.
Neither the computational cost nor the communication cost of the proposed FDIR algorithm (at a given agent) scales with the size of the graph, $|\mathcal V|$, making it an efficient solution for FDIR in large-scale multi-agent systems.
%
%
\vspace{2pt}
\begin{remark}[Warm Start of the ADMM Loop]
In Alg. \ref{alg:admm}, the dual variables $\lbrace \boldsymbol{\lambda}_i^{(l)}\rbrace _{l\in\mathcal E_i}$ and $\lbrace \boldsymbol{\mu}_i^{(j)}\rbrace_{j\in\mathcal N_i}$ are retained between successive SCP iterations (as opposed to re-initializing these variables as $\mathbf 0$), which is referred to as a \textit{warm start} of the ADMM loop.
Using the interpretation of the dual variables given in \cite[Prop. 3.2.2.]{bertsekas}, it can be seen that the optimal values of the dual variables are not expected to change by much after re-linearization of the constraint, thereby explaining why the warm start should yield better convergence properties.
\end{remark}
\vspace{2pt}
\begin{algorithm}[tb]
\caption{Distributed Multi-Agent FDIR (at agent $i$)}
\begin{algorithmic}[1]
\vspace{2pt}
\Require (Global information) $\rho$, $N_{\textrm{SCP}}$, and $N_{\textrm{ADMM}}$; \Statex
\hspace{-20pt} (local information) $\lbrace \mathbf y[l] \rbrace_{l\in \mathcal E_i}$ and $\lbrace \hat {\mathbf p}[j] \rbrace_{j\in \mathcal N_i}$. \vspace{3pt}
\State Initialize the following:
\begin{align}
    \boldsymbol \lambda_{i}^{(l)},\ \boldsymbol \mu_i^{(j)} \ &\leftarrow \mathbf 0 \quad \forall l\in \mathcal E_i, \ \forall j\in \mathcal N_i\\
    \bigstrut[t]\mathbf x^*[j] &\leftarrow \mathbf 0 \quad \forall j\in \mathcal N'_i
\end{align}
where $\mathbf x^*[j]\in\mathbb R^{n_j}$ and $\mathcal N'_i \coloneqq \mathcal N_i \cup \lbrace i \rbrace$.
%
\For{$N_{\textrm{SCP}}$ iterations,} 
\State Linearize the constraint, $\forall j\in\mathcal N'_i,\ l\in\mathcal E_i$:
\begin{align}
\mathbf z[l]&=\mathbf y[l]-\mathbf \Phi\big(\hat {\mathbf p} + \mathbf x^*\big)[l]\\
\mathbf R[l,j]&=\mathbf J_{\mathbf \Phi}\big(\hat{\mathbf p}+\mathbf x^*\big)[l,j]
\end{align}
\State Compute $\lbrace \bar{\mathbf x}[j]\rbrace_{\forall j\in \mathcal N'_i}$ by using the ADMM sub-\Statex \hspace{\algorithmicindent}routine (Alg. \ref{alg:inner}) to solve the linearized sub-problem.
\State Update the error vectors, $\forall j\in\mathcal N'_i$: 
\begin{equation}
\mathbf x^{*}[j]^{^+}\leftarrow \mathbf x^{*}[j]+ \bar{\mathbf x}[j]
\end{equation}
\EndFor
\State \Return The reconstructed error vector at agent $i$, $\mathbf x^*[i]$.
\end{algorithmic}
\label{alg:admm}
\end{algorithm}
\begin{algorithm}[tb]
    \caption{ADMM Subroutine (at agent $i$)}
    \label{alg:inner}
    \begin{algorithmic}[1]
    
    \State Initialize $\bar {\mathbf w}_i^{(j)}\in\mathbb R^{n_i}$ and $\bar {\mathbf w}_j^{(i)}\in\mathbb R^{n_j}$ as $\mathbf 0$, $\forall j\in\mathcal N_i$. 
\vspace{3pt}
\For{$N_{\textrm{ADMM}}$ iterations,}
\State Let $\bar {\mathbf x}[i]$ be the minimizer of
\begin{align*}
\qquad \qquad &\underset{\ 
 \scalebox{0.9}{$\hat{\mathbf x}[i]$}
 }{\textrm{minimize}}
    \ 
    \bigl\|\mathbf x^*[i]+\hat{\mathbf x}[i]\bigr\|
    \\
    & +\sum_{l\in\mathcal E_i} 
    \left(
        \frac{\rho}{2}\bigl\|\mathbf c_i^{(l)}(\hat {\mathbf x}, \bar{\mathbf w})\bigr\|^2 + \boldsymbol{\lambda}_i^{(l)
        \raisebox{2pt}{$\scriptstyle\top$}
        }
        \mathbf c_i^{(l)} (\hat {\mathbf x}, \bar{\mathbf w})
    \right)
    \\
    &+
    \sum_{j\in\mathcal N_i} 
    \left( \frac{\rho}{2}\bigl\|
        \mathbf d_i^{(j)}(\hat {\mathbf x}, \bar{\mathbf w})
        \bigr\|^2 + \boldsymbol{\mu}_i^{(j)
        \raisebox{2pt}{$\scriptstyle\top$}
        }
        \mathbf d_i^{(j)}(\hat {\mathbf x}, \bar{\mathbf w})
    \right)
\end{align*}
\State Communicate $\bar {\mathbf x}[i]$ and $\boldsymbol{\mu}_i^{(j)}$ to agent $j$, $\forall j \in \mathcal N_i$.
\State Let $\lbrace \bar {\mathbf w}_j^{(i)}  |\hsp j\in\mathcal N_i\rbrace$ be the minimizer of
\begin{align*}
 \qquad \qquad &\underset{\ 
 \scalebox{0.85}{$\lbrace\hat{\mathbf w}_j^{(i)}
 | \hsp j\in \mathcal N_i \rbrace$}}{\textrm{minimize}}
    \ 
    \bigl\|\mathbf x^*[i]+\bar{\mathbf x}[i]\bigr\|
    \\
    & +\sum_{l\in\mathcal E_i} 
    \left(
        \frac{\rho}{2}\bigl\|\mathbf c_i^{(l)}(\bar {\mathbf x}, \hat{\mathbf w})\bigr\|^2 + \boldsymbol{\lambda}_i^{(l)
        \raisebox{2pt}{$\scriptstyle\top$}
        }
        \mathbf c_i^{(l)} (\bar {\mathbf x}, \hat{\mathbf w})
    \right)
    \\
    &+
    \sum_{j\in\mathcal N_i} 
    \left( \frac{\rho}{2}\bigl\|
        \mathbf d_j^{(i)}(\bar {\mathbf x}, \hat{\mathbf w})
        \bigr\|^2 + \boldsymbol{\mu}_j^{(i)
        \raisebox{2pt}{$\scriptstyle\top$}
        }
        \mathbf d_j^{(i)}(\bar {\mathbf x}, \hat{\mathbf w})
    \right)
\end{align*}
\State Communicate $\bar{\mathbf w}_j^{(i)}$ to agent $j$, $\forall j \in \mathcal N_i$.
\State Update the dual variables, $\forall j\in\mathcal N_i, l\in\mathcal E_i$:
\begin{align}
\boldsymbol \lambda_i^{(l)^+} &\leftarrow \boldsymbol \lambda_i^{(l)} + \rho \hsp \mathbf c_i^{(l)}(\bar {\mathbf x}, \bar{\mathbf w})\\
\boldsymbol \mu_i^{(j)^+} &\leftarrow \boldsymbol \mu_i^{(j)} + \rho \hsp \mathbf d_i^{(j)}(\bar {\mathbf x}, \bar{\mathbf w})
\end{align}
\EndFor \State \Return An approximate solution to the linearized sub-problem, $\lbrace \bar{\mathbf x}[j] \rbrace_{j\in\mathcal N'_i}$.
    \end{algorithmic}
\end{algorithm}
\subsection{Further Analysis of Algorithm \ref{alg:admm}}
\label{subsec:interpretation}
Consider the term $\|\mathbf x^* + \hat{\mathbf x}\|_{2,1}$ in the objective function of Problem P3. The reason why this term promotes the block sparsity of the error vector is that it is non-differentiable at the points where any of its blocks are equal $\bold 0$. 
Intuitively, if one attempts to `touch' the graph of a non-differentiable (but essentially differentiable) function from below with a set of arbitrarily chosen hyperplanes, the hyperplanes are more likely to touch the non-differentiable points of the graph than not. 
We can make this idea mathematically precise using the concept of subgradients of convex functions.

\vspace{2pt}
\begin{lemma}
Given $\mathbf A\in\mathbb R^{o\times n}$ and $\mathbf b\in\mathbb R^{o}$, where $o$ is a positive integer, let $f:\mathbb R^n \rightarrow \mathbb R$ denote the function
$f(\bold v) = \|\bold v\| + \frac{1}{2}\|\bold A \bold v - \bold b\|^2$.
A vector $\bold v^*\in\mathbb R^n$ is the unique global minimizer of $f$ if and only if either
\begin{itemize}
    \item
$\|\bold A^\top \bold b\| \leq 1 $ and $\bold v^* = \bold 0$, or
\item $\|\bold A^\top \bold b\| > 1 $, $\bold v^*\neq \bold 0$, and
$
\bold A^\top \bold A \bold v^* + \dfrac{\bold v^*}{\|\bold v^*\|}  = \bold A^\top \bold b.
$
\end{itemize}
\vspace{3pt}
\label{lem:proximal}
\end{lemma}
\begin{proof}
The proof is given in Appendix 1.
\end{proof}
%
\begin{theorem}[Thresholding Property]
The minimizer $\bar {\mathbf x}[i]$ of the first primal minimization problem at agent $i$ (step 3 of Alg. \ref{alg:inner}) is given by $\bar{\mathbf x}[i]=-\mathbf x^*[i]$ if and only if
\begin{align}
\Big\|
&\sum_{l \in \mathcal E_i}\mathbf R[l,i]^\top 
\Big(
\mathbf c_i^{(l)}(-\mathbf x^*, \bar{\mathbf w}) + \tilde{\boldsymbol \lambda}_i^{(l)}
\Big) 
\nonumber\\&\qquad \qquad 
+ \sum_{j\in\mathcal N_i}\Big(
\mathbf d_i^{(j)}(-\mathbf x^*, \bar{\mathbf w})
+\tilde{\boldsymbol{\mu}}_i^{(j)}\Big)
\Big\| \leq \frac{1}{\rho},
\label{eq:threshold}
\end{align}
where $\tilde{\boldsymbol{\lambda}}_i^{(l)}\coloneqq \frac{{\boldsymbol{\lambda}}_i^{(l)}}{\rho}$ and $\tilde{\boldsymbol{\mu}}_i^{(j)}\coloneqq \frac{{\boldsymbol{\mu}}_i^{(j)}}{\rho}$.
\label{thm:threshold}
\end{theorem}
\begin{proof}
 Define $\mathbf v_i=\mathbf x^*[i]+\hat{\mathbf x}[i]$, where $\hat{\mathbf x}[i]$ is the variable being minimized over in step 3 of Alg. \ref{alg:inner}. Then, the objective function in step 3 can be expressed as
$\|\mathbf v_i\| + \frac{1}{2}\|\mathbf A_i \mathbf v - \mathbf b_i\|^2 + \gamma_i$,
 where $\gamma_i \in \mathbb R$ is a constant that is independent of $\mathbf v_i$, and
\begin{align}
\mathbf A_i = \sqrt{\rho} \begin{bmatrix}\vspace{-10pt}\\
\mathbf R[l_1, i]\\
\mathbf R[l_2, i]\\
\vdots \vspace{3pt}\\
\mathbf I\\
\mathbf I\\
\vdots\vspace{2pt}
\end{bmatrix},
&\quad 
\mathbf b_i = -\sqrt{\rho} \begin{bmatrix}
    \mathbf c_i^{(l_1)}(-\mathbf x^*, \bar{\mathbf w}) + \tilde{\boldsymbol{\lambda}}_i^{(l_1)}\\
    \mathbf c_i^{(l_2)}(-\mathbf x^*, \bar{\mathbf w}) + \tilde{\boldsymbol{\lambda}}_i^{(l_2)}\\
    \vdots \\
    \mathbf d_{i}^{(j_1)}(-\mathbf x^*, \bar{\mathbf w}) + \tilde{\boldsymbol{\mu}}_{i}^{(j_1)}\vspace{2pt}\\
    \mathbf d_{i}^{(j_2)}(-\mathbf x^*, \bar{\mathbf w}) + \tilde{\boldsymbol{\mu}}_{i}^{(j_2)}\\
    \vdots
\end{bmatrix},
\label{eq:A_i-b_i}
\end{align}
where $l_1,l_2, \dots, l_{|\mathcal E_i|}$ denote the elements of $\mathcal E_i$, and $j_1, j_2, \dots, j_{|\mathcal N_i|}$ denote the elements of $\mathcal N_i$.
Note that the minimization of $\|\mathbf v_i\| + \frac{1}{2}\|\mathbf A_i \mathbf v_i - \mathbf b_i\|^2$ over $\mathbf v_i$ is equivalent to the minimization problem in step 3 of Alg. \ref{alg:inner}, as neither the constant factor $\gamma_i$ nor the change of variables $\hat{\mathbf x}[i]\rightarrow \mathbf v_i$ changes the underlying optimization problem.
Thus, we can use Lemma \ref{lem:proximal} to complete the proof.
\end{proof}

Theorem \ref{thm:threshold} can be used to obtain a meaningful interpretation of the proposed multi-agent FDIR algorithm, as follows. Firstly, 
observe that agent $i$ sets its error vector to $\mathbf 0$ if and only if the value on the left-hand side of inequality (\ref{eq:threshold}), which may be interpreted as the \textit{residual} of agent $i$, lies below a certain \textit{threshold}. The residual is a function of the vectors in $\lbrace \bar{\mathbf w}_i^{(j)} \rbrace_{j\in\mathcal N_i}$, which are updated by the neighbors of agent $i$. This interdependency between the agents' residuals is what enables the proposed algorithm to identify the faults in a collaborative manner.
Furthermore, we see that the penalty parameter $\rho$ is inversely proportional to the threshold in inequality (\ref{eq:threshold}). Lastly, observe that step 7 of Algorithm \ref{alg:inner} can be rewritten as: $\tilde{\boldsymbol \lambda}_i^{(l)+} \leftarrow \tilde{\boldsymbol \lambda}_i^{(l)} + \hsp \mathbf c_i^{(l)}(\bar {\mathbf x}, \bar{\mathbf w})$, which means that $\tilde{\boldsymbol \lambda}_i^{(l)}$ (and similarly, $\tilde{\boldsymbol \mu}_i^{(j)}$) is a running sum of the amount of constraint violation in each iteration. Hence, a repeated violation of the constraints at agent $i$ compels the agent to set its error vector to a non-zero value, whereas the term $\mathbf c_i^{(l)}(-\mathbf x^*, \bar{\mathbf w})$ in the residual checks whether this constraint violation can be attributed to the neighbors of agent $i$ instead.

A secondary contribution of Theorem \ref{thm:threshold} is that it informs an efficient strategy for minimizing the non-differentiable function in step 3 of Algorithm \ref{alg:inner}.
In general, the minimization of a non-differentiable function using subgradient descent achieves a convergence speed of $O(1/\sqrt{t})$ \cite{gordon2012subgradient}. However, if the inequality in (\ref{eq:threshold}) holds at agent $i$, then $\bar{\mathbf x}[i]$ can be set to $-\mathbf x^*[i]$, bypassing the need to solve an optimization problem altogether. If the inequality does not hold, then Lemma \ref{lem:proximal} guarantees that the minimizer is bounded away from the non-differentiable point. Therefore, one can use Nesterov's accelerated gradient descent method
to achieve a convergence speed of $O(1/t^2)$ \cite{nesterov1983method}, as long as the gradient descent procedure is initialized within a neighborhood of the minimizer where the function is differentiable.

%
%
\section{Numerical Simulations}
\label{sec:numerical}
In this section, we consider the example of a multi-agent system consisting of unmanned aerial vehicles (UAVs) that are able to measure their distances from each other (which is the same scenario that was considered in the example in Section \ref{subsec:example}). In practice, the inter-agent distances can be measured using the time-of-arrival (ToA) of inter-agent communications \cite{wen2020multi}.
Each agent is also equipped with a GNSS receiver which enables it to estimate its own position.
However, some of the agents (selected at random) have estimation errors due to multipath effects and/or loss of GNSS tracking loops, which are two types of GNSS faults that are known to occur in practice \cite{siebert2022multipath}. Using numerical simulations, it is demonstrated that the proposed distributed multi-agent FDIR algorithm (Algs. \ref{alg:admm} and \ref{alg:inner}) can process the inter-agent distances to correctly identify the faulty agents and reconstruct their error vectors. To keep within the scope of this paper, it is assumed that there is no noise in the inter-agent distance measurements\footnote{The effect of measurement noise was investigated, both analytically and numerically, in \cite{khan2023recovery}.}.

\subsubsection*{Simulation Scenario} 
\begin{figure}
         \centering
         \includegraphics[trim={0.7cm, 1.5cm, 0.8cm, 2.0cm},clip, width=0.9\columnwidth]{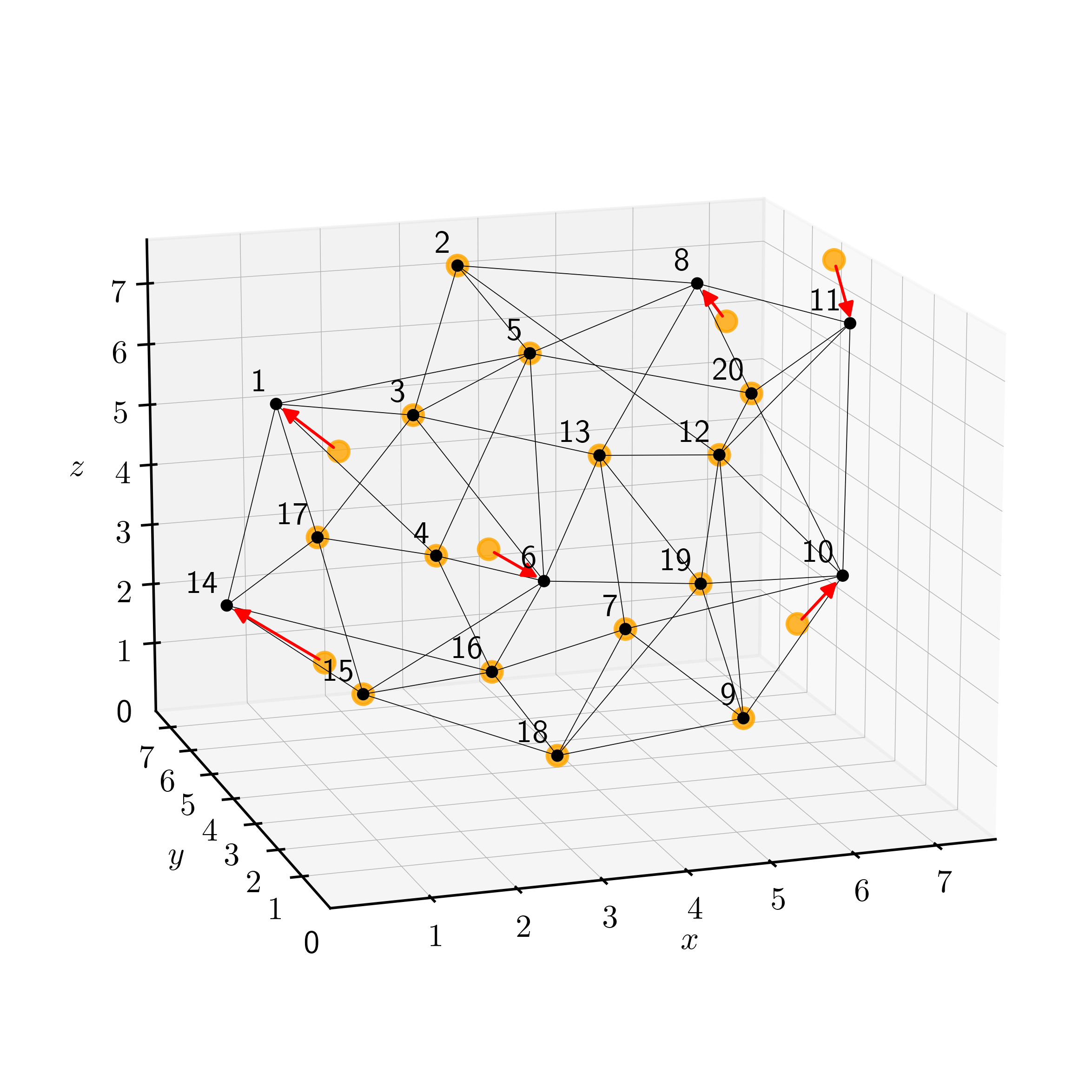}
        \caption{The black dots correspond to the true states of the multi-agent system, whereas the yellow discs represent the agents' estimated states.
        The red arrows depict the non-zero blocks of the reconstructed error vector after $60$ iterations (in total) of the ADMM loop.}
        \label{fig:single-run}
\end{figure}

The parameters of the simulation are as follows. The multi-UAV configuration $(\mathcal G, \mathbf x)$ consists of $20$ UAVs, where $\lbrace\mathbf p[i] \in \mathbb R^3\rbrace_{i\in\mathcal V}$ represent their position vectors, depicted by the black dots in Fig. \ref{fig:single-run}. In the figure, the black lines are used to connect the pairs of agents that are able to measure their distances from each other. 
For the configuration in Fig. \ref{fig:single-run}, the rank of $\mathbf J_{\mathbf \Phi}(\mathbf p)$ was determined to be $54$; its first six eigenvalues are close to $0$ (smaller than $10^{-13}$ in magnitude) and its seventh-smallest eigenvalue is non-zero (on the order of $10^{-4}$). Moreover, we have $n=3|\mathcal V|=60$ as the dimension of the configuration space.
Hence (using Theorem \ref{thm:generalized}), the search-space of the error reconstruction problem is a $6$-dimensional embedded submanifold in $\mathbb R^{60}$.

In each simulation, the set of faulty agents, $\mathcal D \subseteq \mathcal V$, is constructed by selecting $6$ agents at random. Given a faulty agent $i\in\mathcal D$, we let its position estimate be $\hat{\mathbf p}[i] = \mathbf p[i] + \mathbf x[i]$, where $\mathbf x[i]\in\mathbb R^3$ is sampled from a random vector having a uniform distribution over the domain $[-1,1]^3$ (i.e., each element of $\mathbf x[i]$ lies between $[-1,1]$). Given a fault-free agent $i\in\mathcal D^\complement$, $\mathbf x[i] $ is set to $\mathbf 0$.
The position estimates of the agents are depicted by the yellow discs in Fig. \ref{fig:single-run}.

\subsubsection*{Multi-Agent FDIR Performance} The proposed distributed multi-agent FDIR algorithm is implemented using the following parameters. We set $\rho=1$ and $N_{\textrm{ADMM}}=10$, i.e., the nonlinear constraint is re-linearized after every $10$ ADMM iterations. Figure \ref{fig:single-run} shows the reconstructed error vector $\mathbf x^*$ after $6$ SCP iterations (which corresponds to a total of $60$ ADMM iterations). Only the blocks of $\mathbf x^*$ whose Euclidean norm is larger than $10^{-5}$ are visualized in this figure; these `non-zero' blocks are seen to correspond precisely to the faulty agents. 

\begin{figure}
         \centering
         \includegraphics[width=0.49\textwidth]{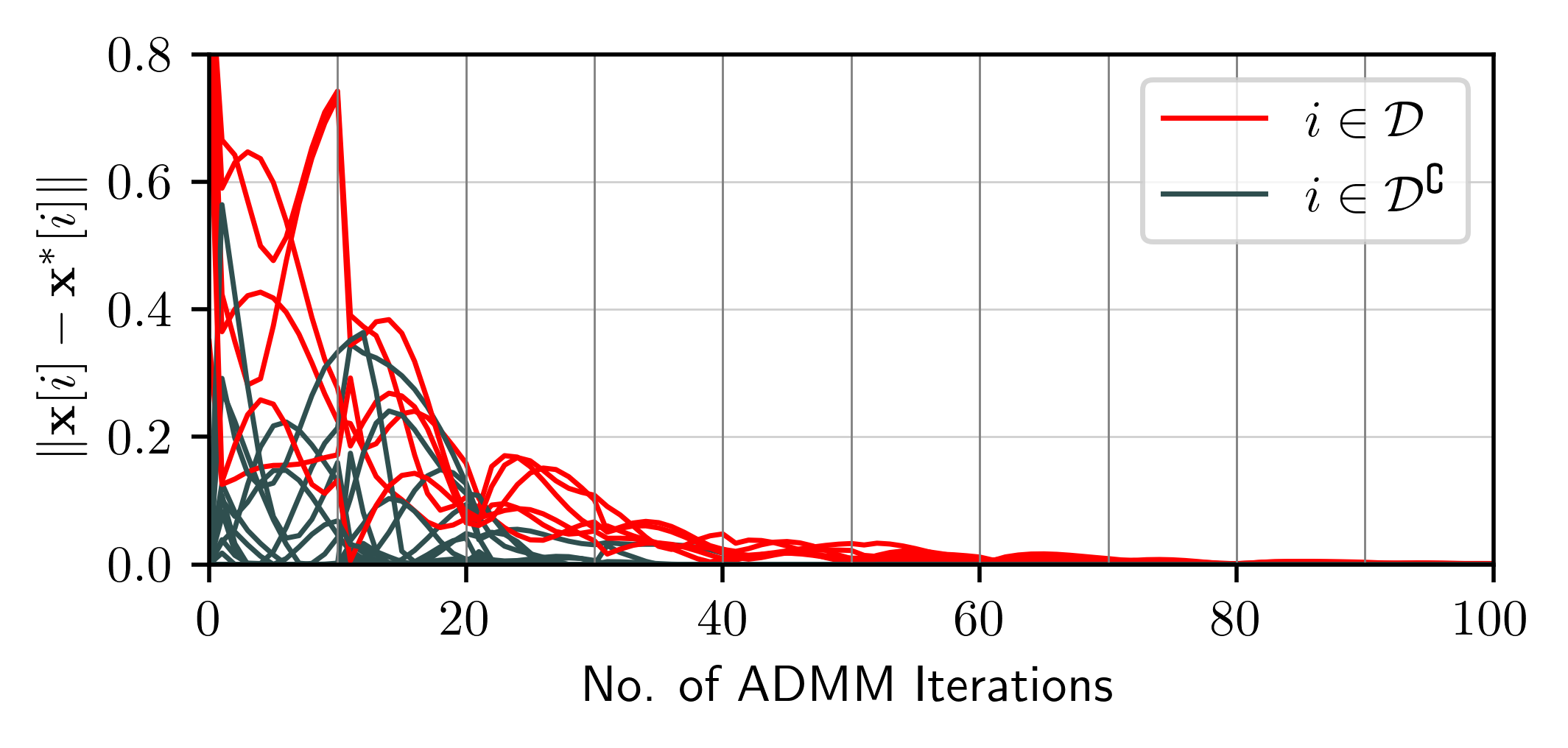}
        \caption{The accuracy of the reconstructed error vector at each agent. The red lines correspond to the agents in $\mathcal D$ (i.e., the faulty agents), and the vertical grey lines indicate the iterations where the constraint was re-linearized.}
        \label{fig:block_errors}
\end{figure}

\begin{figure}[h]
         \centering
         \begin{subfigure}[t]{\linewidth}
         \centering
         \includegraphics[width=\linewidth]{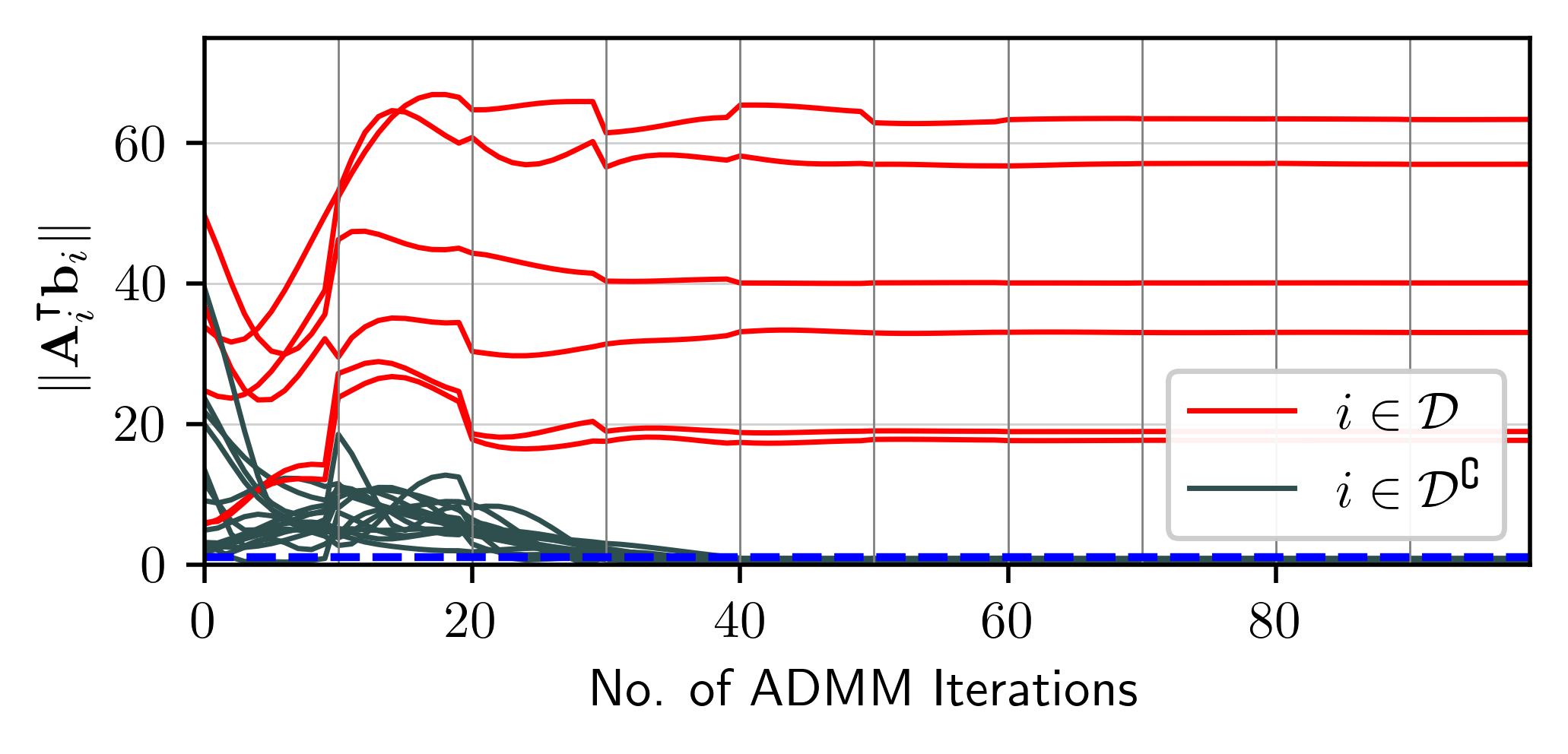}
         \caption{$\rho = 1.0$}
         \label{fig:thresholds}
         \end{subfigure}
         \begin{subfigure}[t]{\linewidth}
         \centering
         \includegraphics[width=\linewidth]{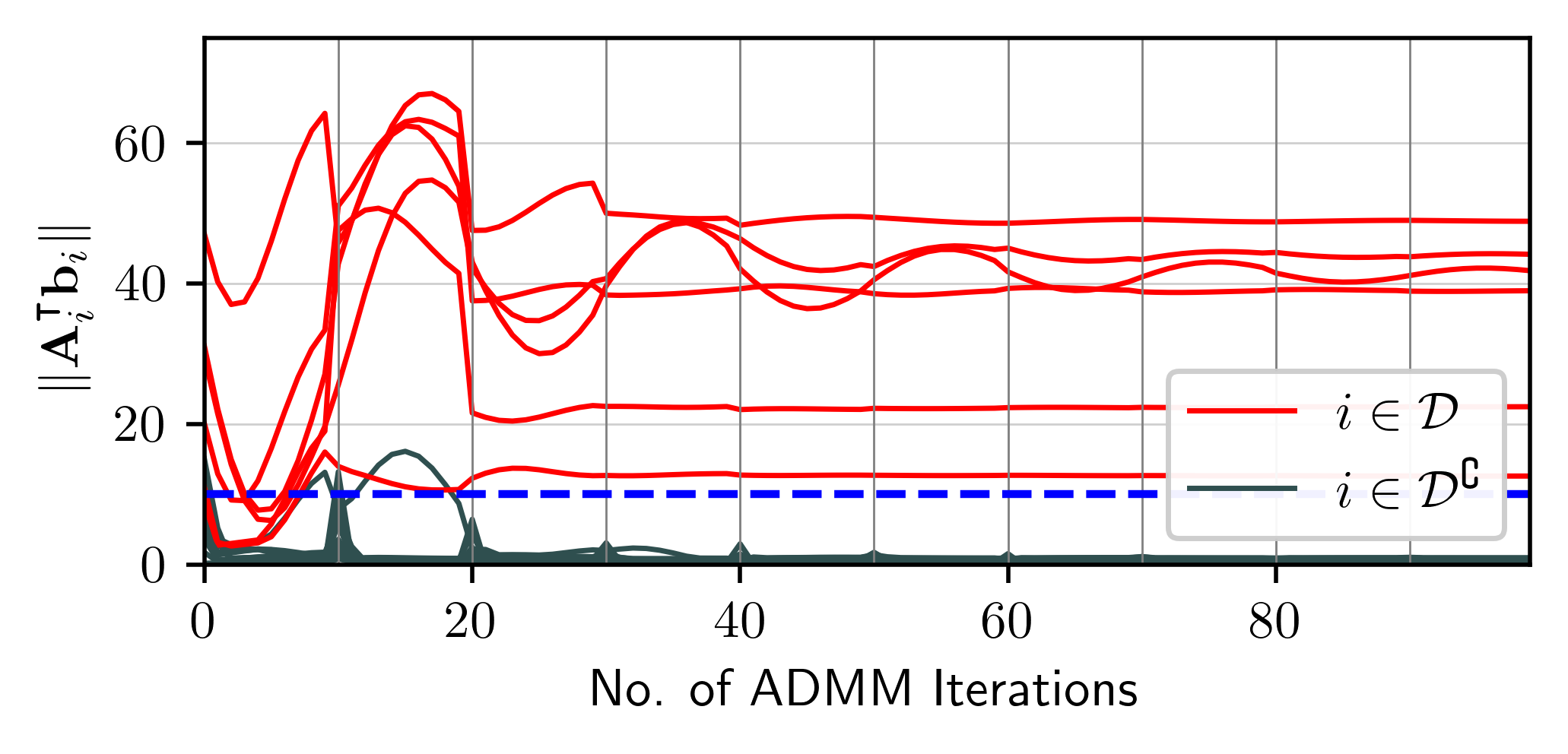}
         \caption{$\rho = 0.1$}
         \label{fig:thresholds_2}
         \end{subfigure}
        \caption{The fault-detection threshold $1/\rho$ (blue dashed line) and the residual energy $\|\mathbf A_i^\top \mathbf b_i\|$ are visualized for each agent; the definitions of $\mathbf A_i$ and $\mathbf b_i$ are given in (\ref{eq:A_i-b_i}). The red lines correspond to the agents in $\mathcal D$ (i.e., the faulty agents), and the vertical grey lines indicate the iterations where the constraint was re-linearized.}
\end{figure}

Figure \ref{fig:block_errors} visualizes the accuracy of the reconstructed error vector at each agent, which is represented by the quantity $\|\mathbf x[i] - \mathbf x^*[i]\|$; the red lines correspond to the agents in $\mathcal D$, i.e., the faulty agents. It can be observed that the error at each agent asymptotically tends to $0$. Furthermore, a defining feature of the proposed multi-agent FDIR algorithm is that it is interpretable; specifically, we showed in Theorem \ref{thm:threshold} that 
each agent does something akin to \textit{residual testing} in order to determine whether it has a fault. Figure \ref{fig:thresholds} depicts the residuals of each agent, as well as the fault-detection threshold (blue dashed line), whose expressions can be found in Theorem \ref{thm:threshold}. As expected, the faulty agents are precisely the ones whose residuals have an energy greater than the detection threshold. Figure \ref{fig:thresholds_2} depicts the same plot for $\rho = 0.1$, confirming our observation in Theorem \ref{thm:threshold} that the penalty parameter is inversely proportional to the fault-detection threshold. Furthermore, observe that the faults are correctly identified within $40$ and $20$ ADMM iterations in Figures \ref{fig:thresholds} and \ref{fig:thresholds_2}, respectively, even though the error reconstruction mechanism takes longer (about $80$ ADMM iterations) to converge (cf. Fig. \ref{fig:block_errors}). This is consistent with the existing literature on ADMM, which notes that the dual variables of ADMM converge rapidly to near-optimal values, and asymptotically to their optimal values.
On the other hand, convergence of the primal variables of ADMM is not guaranteed in general (see Sec. 3.2.1 of \cite{boyd2011distributed}). 

\subsubsection*{Monte Carlo Simulations} 
\begin{figure}
         \centering
         \includegraphics[width=0.49\textwidth]{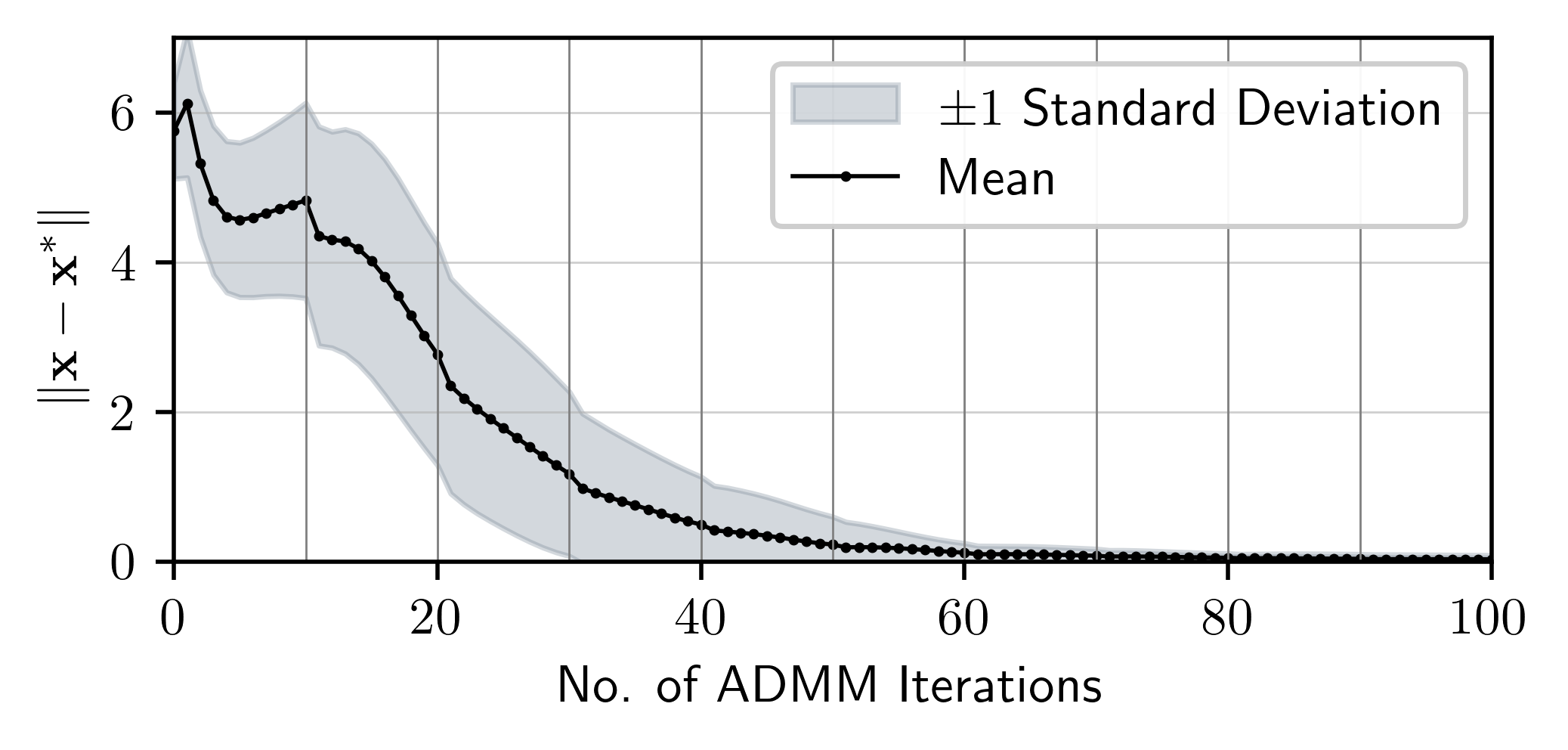}
        \caption{The accuracy of the error reconstruction mechanism is visualized in terms of the mean and $\pm 1$ standard deviation of $\|{\mathbf x} - \mathbf x^*\|$ (depicted by the black line and the shaded region, respectively) across $100$ Monte Carlo simulations. The vertical grey lines indicate the iterations where the constraint was re-linearized.}
        \label{fig:all_errors}
\end{figure}

To further investigate the accuracy of the error reconstruction mechanism, the numerical simulation was repeated for $100$ Monte Carlo (MC) simulations. We select $6$ faulty agents and execute $10$ SCP iterations in each MC simulation, such that the identities of the faulty agents and their error vectors are randomly sampled at the beginning of each MC simulation. Figure \ref{fig:all_errors} visualizes the quantity $\|\mathbf x - \mathbf x^*\|$, which represents the accuracy of the reconstructed multi-agent error vector; the black line represents its mean and the shaded region corresponds to $\pm 1$ standard deviation. It can be seen that the reconstructed error vector $\mathbf x^*$ approximates the true error vector $\mathbf x$ quite well and that the standard deviation of the quantity $\|\mathbf x - \mathbf x^*\|$ tends to $0$. Thus, the proposed multi-agent FDIR algorithm is able to accurately reconstruct the multi-agent error vector with high probability, irrespective of the identities of the faulty agents.

%

\section{Conclusions}
\label{sec:conclusions}
In this work, we studied the problem of fault detection, identification and reconstruction (FDIR) of multi-agent systems, which presents distinctive challenges due to the combinatorial complexity of the conventional FDIR algorithms.
The proposed approach for multi-agent FDIR borrows ideas from compressive sensing to reformulate the FDIR problem as one of reconstructing a block-sparse error vector. The search-space of the error-reconstruction problem was characterized and its connections to the exiting literature on rigidity theory were explored, which explains why the proposed approach is able to reconstruct the error vector without requiring the identities of the anchors (i.e., the agents that do not have faults) to be known \textit{a priori}. Thereafter, we developed a distributed algorithm that solves the error-reconstruction problem at each agent. Numerical simulations were used to validate the effectiveness of the proposed distributed multi-agent FDIR algorithm.

One possibility for future work is to consider each agent to evolve according to a dynamical model, similar to \cite{kazumune2020distributed}. Another future research direction is to cast the multi-agent FDIR problem in the Bayesian setting.
\appendix[Proof of Lemma \ref{lem:proximal}]
\begin{proof}
Using the property of subdifferentials \cite[Thm. 4.1.1]{hiriart2004fundamentals}, we have
\begin{equation}
    \partial f(\bold v) = \partial g(\bold v) + \lbrace \bold A^\top \bold A \bold v - \bold A^\top \bold b\rbrace
\end{equation}
where $\partial f(\bold v)$ is the subdifferential (i.e., the set of subgradients) of $f$ at $\bold v$, $g(\bold v)=\|\bold v\|$, and `$+$' denotes the Minkowski sum. $g(\bold v)$ is differentiable everywhere but at the origin.
The subdifferential of $g(\bold v)$, where $\bold v\in \mathbb R^n$, is
\begin{align}
    \partial g({\bold v}) = \begin{cases}
    \begin{array}{cc}
        \left\lbrace \dfrac{{\bold v}}{\|{\bold v}\|} \right\rbrace & {\bold v} \neq \bold 0 \vspace{5pt}\\
        \left\lbrace \bold u \hspace{1pt}\big|\hspace{1pt} \bold u\in \mathbb R^n, \|\bold u\|\leq 1 \right\rbrace & {\bold v} = \bold 0
    \end{array}
    \end{cases}
\end{align}
\textbf{Case $\|\bold A^\top \bold b\| \leq 1$:}
By choosing the subgradient $\bold A^\top \bold b$ at $\bold v = \bold 0$, we see that $\bold 0 \in \partial f(\bold 0)$.  Moreover, if $\bold b = \bold 0$ then $\bold 0$ is clearly a unique global minimum. Now let $\bold b\neq \bold 0$. To prove that $\bold 0$ is the unique point where a subgradient of $f$ vanishes, we use the fact that $\|{}\cdot{}\|$ is only `flat' in the radial directions, whereas the hyperplane corresponding to $\bold A \bold v = \bold b$ is an affine subspace which does not contain any of the radial lines. To formalize this, assume there is another point $\tilde {\bold v}$ such that $\bold 0\in \partial f(\tilde{\bold v})$. Since $f(\bold 0)  \geq f(\tilde{\bold v}) $ and $f(\tilde{\bold v})\geq f(\bold 0) $, $f(\bold 0) = f(\tilde{\bold v})$. Choose a number $\theta$ such that $0\leq \theta \leq 1$.
From convexity, $f(\theta \tilde{\bold v})\leq f(\bold 0)$, and because $\bold 0 \in \partial f(\bold 0)$, $f(\theta \tilde{\bold v})\geq f(\bold 0) $. Thus, for all $0\leq \theta \leq 1$, we have $f(\bold 0) = f(\theta \tilde{\bold v}) = f(\tilde {\bold v})$, which implies that $\tilde {\bold v} \in \ker(\nabla^2 f(\theta \tilde {\bold v}))$. Therefore, we have
\begin{equation}
\tilde{\bold v} \in \ker\left( \frac{1}{\theta\|\tilde {\bold v}\|}\left(I - \frac{\tilde {\bold v} \tilde {\bold v}^\top}{\|\tilde {\bold v}\|^2}\right) + A^\top A\right)
\end{equation}
or equivalently, $\tilde {\bold v} \in \ker (\bold A)$. Setting the gradient at $(\theta \tilde {\bold v})$ to $\bold 0$, we get
\begin{equation}
    \frac{\tilde{\bold v}}{\|\tilde{\bold v}\|} = \bold A^\top \bold b
\end{equation}
but a vector cannot be in both the kernel of $\bold A$ and the range space of $\bold A^\top$, giving us a contradiction.

\noindent\textbf{Case $\|\bold A^\top \bold b\| > 1$:}
If $\bold v^*\neq0$, then $\partial f(\bold v^*)$ is a singleton containing the gradient vector, and both directions of the claim follow from setting the gradient to $\bold 0$. The uniqueness of $\bold v^*$ follows from similar arguments to the ones we used above.

To see that $\bold v^* \neq \bold 0$, choose a unit vector $\bold u\in \mathbb R^n$ and consider that
\begin{align}
 \lim_{k\rightarrow 0^+}\nabla f(k \bold u) &= \lim_{k \rightarrow 0^+} k \bold A^\top \bold A \bold u + \bold u - \bold A^\top \bold b\\
&=  \bold u - \bold A^\top \bold b
\end{align}
where $\nabla f(\hsp \cdot \hsp)$ is the gradient of $f$, which is well-defined at all non-zero points. Since $\|\bold A^\top \bold b\|>1$, we can always choose a direction $\bold u$ such that $f(\bold v)$ is strictly decreasing along that direction near the origin, which completes the proof.
\end{proof}

\bibliographystyle{ieeetr}
\bibliography{IEEEabrv,refs}
\begin{IEEEbiographynophoto}
{Shiraz Khan} received his MS degree in Aeronautics and Astronautics from Purdue University in 2020, and his Bachelor's degree in Aerospace Engineering from IIT Madras in 2018. He is currently pursuing a Ph.D. degree at Purdue University, as part of which he investigates the problems of robustness, resilience, and scalability of state estimation. The potential applications of his Ph.D. research include decentralized and cyberattack-resilient state estimation in single and multi-agent scenarios.
One of his current research interests is in exploring and leveraging the underlying structure (such as sparsity, geometry, and/or symmetry) of various control theory and signal processing applications.
\end{IEEEbiographynophoto}

\begin{IEEEbiographynophoto}
{Inseok Hwang} received the Ph.D. degree in Aeronautics and Astronautics from Stanford University, and is currently a professor in the School of Aeronautics and Astronautics and a university faculty scholar at Purdue University. His research interests lie in high assurance autonomy (which guarantees safety, security and performance) for Cyber-Physical Systems (CPS) based on the hybrid systems approach and its applications to safety critical systems such as aircraft/spacecraft/unmanned aerial systems (UAS), air traffic management, and multi-agent systems, which are complex systems with interacting cyber (or logical or human) elements and physical components. For his research, he leads the Flight Dynamics and Control/Hybrid Systems Laboratory at Purdue University. He received various awards, including the National Science Foundation (NSF) CAREER award in 2008, recognition as one of the nation’s brightest young engineers by the National Academy of Engineering (NAE) in 2008, the AIAA Special Service Citation in 2010, the University Faculty Scholar by Purdue university in recognition of outstanding scholarship in 2017, Outstanding Graduate Faculty Mentor award 2018, and the C.T. Sun Award in recognition of excellence in research in 2019. He is an Associate Fellow of AIAA and a member of IEEE Control Systems Society and Aerospace and Electronic Systems Society. He is currently an associate editor of the IEEE Transactions on Aerospace and Electronic Systems.
\end{IEEEbiographynophoto}

\end{document}